\documentclass[floats,prb,twocolumn,amsmath,superscriptaddress]{revtex4}

\pdfoutput=1

\usepackage{graphicx}
\usepackage{color}
\usepackage[vertfit]{breakurl}

\usepackage{dcolumn}

\begin{document}

\renewcommand{\rm} [1]{\mathrm{#1}}
\renewcommand{\bf} [1]{\textbf{#1}}
\renewcommand{\cal} [1]{\mathcal{#1}}

\newcommand{\ket}    [1]{{|#1\rangle}}
\newcommand{\bra}    [1]{{\langle#1|}}
\newcommand{\braket} [2]{{\langle#1|#2\rangle}}
\newcommand{\bracket}[3]{{\langle#1|#2|#3\rangle}}

\newcommand{\red}{\textcolor{red}}
\newcommand{\blue}{\textcolor{blue}}
\newcommand{\green}{\textcolor{green}}
\newcommand{\cyan}{\textcolor{cyan}}
\newcommand{\magenta}{\textcolor{magenta}}

\def\numkpts{N_0}
\def\numbands{N}

\def\pw{^{({\rm W})}}
\def\ph{^{({\rm H})}}
\def\k{{\textbf k}}
\def\R{{\textbf R}}
\def\G{{\textbf G}}
\def\r{{\textbf r}}
\def\a{{\textbf a}}
\def\b{{\textbf b}}
\def\q{{\textbf q}}
\def\0{{\textbf 0}}
\def\x{{\textbf x}}
\def\s{{\textbf s}}

\def\D{{D}\ph}
\def\n{{\cal N}}
\def\u{{\cal U}}
\def\o{{\cal O}}
\def\e{{\cal E}}
\def\v{{\rm v}}

\def\la{\langle\kern-2.0pt\langle}
\def\ra{\rangle\kern-2.0pt\rangle}
\def\vt{\vert\kern-1.0pt\vert}

\def\AA{\buildrel_\circ\over{\mathrm{A}}}

\renewcommand{\vec}[1]{{\mathbf #1}}

\newcommand{\ie}{\textit{i.e.}\ }
\newcommand{\eg}{\textit{e.g.}\ }

\newcommand\TM{\rule{0pt}{3.8ex}}
\newcommand\T{\rule{0pt}{3.0ex}}
\newcommand\B{\rule[-1.4ex]{0pt}{0pt}}
\newcommand\BM{\rule[-1.8ex]{0pt}{0pt}}
\newcommand\TL{\rule{0pt}{2.7ex}}
\newcommand\BL{\rule[-0.6ex]{0pt}{0pt}}
\newcommand\TMED{\rule{0pt}{2.7ex}}
\newcommand\BMED{\rule[-1.2ex]{0pt}{0pt}}

\title{Evolution of the Fermi surface of arsenic through the rhombohedral to 
simple-cubic phase transition: a Wannier interpolation study}

\author{Patricia K. Silas}
\affiliation{Theory of Condensed Matter, Cavendish Laboratory,
University of Cambridge, JJ Thomson Avenue, Cambridge CB3 0HE, United Kingdom}

\author{Peter D. Haynes}
\affiliation{Departments of Materials and Physics,
Imperial College London, Exhibition Road, London SW7 2AZ, United Kingdom}

\author{Jonathan R. Yates}
\affiliation{Department of Materials,
University of Oxford, Parks Road, Oxford OX1 3PH, United Kingdom}

\date{\today}
\begin{abstract}

The pressure dependence of the Fermi surface of arsenic is examined
using the technique of Wannier interpolation, enabling a
dense sampling of the Brillouin zone and
the ability to capture fine features within it. 
Focusing primarily on the A7~$\to$~simple-cubic
phase transition, we find that this semimetal
to metal transition is accompanied by the folding of Fermi
surfaces. 
The pressure dependence of the density of states (DOS) of
arsenic indicates that the onset of the Peierls-type cubic to
rhombohedral distortion is signified by the appearance of emerging van
Hove singularities in the DOS, especially around the Fermi level. 
As we noted in an
earlier study, high levels of convergence are consequently required
for an accurate description of this transition.

\end{abstract}


\maketitle

\section{Introduction}
\label{sec:intro}
\vspace{-1mm}

At ambient pressures the \mbox{group--V} element arsenic is a semimetal
with the rhombohedral $\alpha$-As or A7 crystal structure (space group \textit{R$\bar{3}$m}). 
Under applied
pressure arsenic undergoes a series of structural phase 
transitions,~\cite{degtyareva_04_review} the
first of which occurs at approximately 28~GPa yielding the metallic simple-cubic
(sc) structure (space group \textit{Pm$\bar{3}$m}).
The change in the electronic structure of arsenic across this semimetal
to metal phase transition is described most naturally using the 
Fermi surface.
In the A7 phase, arsenic is known to have a very unusual
Fermi surface, consisting of six lobes connected by six long thin
cylinders or ``necks''.  This Fermi surface is today still
depicted as it was by Lin and Falicov~\cite{linandfalicov_66} in
1966---an artist's rendition---it is a famous image that is often
used to describe the complexity that a Fermi surface may exhibit.
First-principles calculations on the very dense grids
required to resolve the fine details of the Fermi surface of arsenic
are computationally expensive even with contemporary computing resources.
Here we demonstrate that the method of Wannier
interpolation~\cite{yates_07,x_wang_06,x_wang_07,giustino_07}
achieves the same accuracy far more inexpensively.

In this study we use Wannier interpolation
to investigate the pressure dependence of the Fermi surface
and density of states (DOS) of arsenic, in particular as 
it experiences the A7~$\to$~sc semimetal to metal phase transition.
Wannier interpolation of the band structure is
 based on the use of maximally-localized
Wannier functions (MLWFs),~\cite{marzari_97,souza_01,marzari_12}
which act in the spirit of ``tight-binding'' as a set of spatially-localized
orbitals, allowing calculations to be performed with the accuracy of first
principles simulations but at low computational cost.
Indeed, arsenic provides us with a beautifully simple demonstration
of the power of Wannier interpolation---it is an ideal testing ground
for the use of this technique to study phase transitions
involving a metal.
%


%
In an earlier work we used density-functional theory (DFT)
to determine the transition pressure of the 
A7~$\to$~sc phase transition of arsenic.~\cite{silas_yates_haynes_08}
The investigation consisted of performing relaxations of the two-atom
unit cell subjected to pressures
over the range of \mbox{0--200~GPa}.  The study
was focused primarily on the A7~$\to$~sc
semimetal to metal phase transition.  
That work enabled us to address 
a long-standing question as to whether experimental results 
for the transition pressure obtained by
 \mbox{Beister, \textit{et al.}}~\cite{beister_90}
were correct over those of Kikegawa and Iwasaki,~\cite{kikegawa_87}
with our results supporting the findings of the former.
It further allowed us to explain the wide range of theoretical values
for the transition pressure that can be found in the 
literature---the spread of values is due to the difficulty of adequately
sampling the Brillouin zone (BZ).
We discovered that high-quality results required extensive
convergence testing with respect to BZ sampling
and amount of smearing used.
In particular, we found that in order to ensure convergence
and reliability of results when studying phase transitions
involving a metal, dense BZ sampling grids are essential.~\cite{silas_08}
We thus begin by using Wannier interpolation to determine the 
Fermi surface of arsenic at 0~GPa.
We inspect closely the finest features of the electron
and hole Fermi surfaces at ambient pressures
and make a quantitative comparison of our results
with those available in the literature.
Building on the results obtained from the relaxations
performed in our earlier study,~\cite{silas_yates_haynes_08} we
then use Wannier interpolation to investigate
the pressure dependence of the Fermi surface of arsenic,
and in particular to study exactly how it evolves through the 
A7~$\to$~sc transition.
Finally, we use Wannier interpolation to investigate the DOS of arsenic
over a range of different pressures.
In our earlier study, we found that converging our geometry
optimizations in the vicinity of the A7~$\to$~sc phase transition
was rather more difficult than expected.~\cite{silas_yates_haynes_08,silas_08}
We will see that this is due
to the rapid change of the DOS around the Fermi level across
the transition. Consequently, high levels of convergence are required
in order to calculate the transition pressure accurately.
%


\vspace{1mm}
The paper is organized as follows.  Section~II outlines
important technical aspects of the methodology,
and Sec.~III deals with the computational details of our
calculations.
In Secs. IV--VII, we present and discuss our results:
we reveal the actual Fermi surface of
uncompressed arsenic---the
features thereof are compared to
results of theory and experiment that exist in the literature;
we study the pressure dependence of the Fermi surface
of arsenic, with a view to observing it as it undergoes
the A7~$\to$~sc structural phase transition;
we calculate the Wannier-interpolated DOS 
of arsenic in the A7 and sc phases, and we
present the pressure dependence
of the DOS of arsenic in the immediate vicinity of
the A7~$\to$~sc transition; we conclude with a brief discussion.
%


\section{Methodology}
\label{sec:methodology}

\subsection{Wannier Interpolation}
\label{sec:wannier_interpolation}

%
A conventional electronic-structure calculation yields a set of Bloch
states describing the low-lying band structure of a particular
system in its ground state. A set of (generalized)
 Wannier functions~\cite{wannier_37}
(WFs) can be constructed from these Bloch states
in order to be able to describe the system in real space, with the WFs
acting as a basis set of localized orbitals.  
The method of Wannier interpolation~\cite{yates_07,x_wang_06,x_wang_07,giustino_07}
is based on the use of a \textit{unique} set of Wannier functions, the so-called
``maximally-localized Wannier functions'',~\cite{marzari_97,souza_01,marzari_12}
which serve as an ``exact tight-binding'' basis for the target system.
Prescriptions due to Marzari and Vanderbilt~\cite{marzari_97}
and Souza, Marzari and Vanderbilt~\cite{souza_01} allow the
determination of this unique set of Wannier functions---the
former in the case of a set of isolated bands and the latter
extended to the case of a set of entangled bands---by requiring the
resulting WFs to exhibit minimum spread. 
Wannier interpolation is a simple and widely
applicable scheme that incorporates the idea
of a tight-binding model with the accuracy of \textit{ab initio}
methods.  The technique is ideal for the evaluation
of quantities that require a very dense sampling
of the BZ.  In particular,
it should be very useful for the study of metallic
systems as it enables a sampling of the Fermi surface
to \textit{ab initio} accuracy but with little associated cost.
Here we are concerned with finding the eigenvalues
of the one-electron effective Hamiltonian
at arbitrary $\k$, $\varepsilon_{n\k}$, where $n$ is the band index.
However, this interpolation procedure can be extended to
the calculation of other quantities of interest---the
eigenvalues of any periodic one-electron operator
can be determined in a similar fashion.~\cite{yates_07}

\vspace{-5mm}
\subsection{Adaptive Smearing}
\label{sec:adaptive_smearing}

%
To accelerate the convergence of singularities in the DOS,
we apply a state-dependent broadening scheme as follows.
Within the Wannier interpolation method, we have the ability
to vary the smearing of the occupancies according to the band gradients
 $|\partial \varepsilon_{n\k} / \partial \k|$,
as the gradients can be computed efficiently at no additional cost.
The state-dependent smearing width $W_{n\k}$ is set to~\cite{yates_07}
\begin{equation}
 W_{n\k} = a \left| \frac{\partial \varepsilon_{n\k}}{\partial \k} \right| \Delta k,
\end{equation}
where $a$ is a dimensionless constant, of value 0.2 in this study.
This state-dependent broadening scheme
is termed ``adaptive smearing''~\cite{yates_07} and it
enables highly resolved DOSs, especially when combined
with the method of cold smearing.~\cite{marzari_99}
%

\section{Computational Details}
\label{sec:comp_details}

%
We build on the findings of our earlier studies of arsenic,~\cite{silas_yates_haynes_08}
in which structural relaxations were performed
on the two-atom rhombohedral unit cell subjected
to a range of pressures.
These geometry optimizations resulted in 
a series of pressure-dependent configurations,
characterized by the lattice parameters which are  
described in the discussion of the unit cell of arsenic
found in the material supplementary to this study.~\cite{silas_12}
For each configuration (each pressure),
we repeat exactly the same procedure.
The computational details are as follows.  
For each pressure investigated, we first obtain the Bloch states
self-consistently using the density-functional-based
{\tt PWSCF} code~\cite{pwscf} in the local-density approximation (LDA).
We use a scalar-relativistic, norm-conserving
pseudopotential generated using the ``atomic'' code
(by A. Dal Corso, supplied with the Quantum Espresso
distribution~\cite{pwscf}) to describe the core-valence interaction
of arsenic.  The pseudopotential core radii are as
follows: 1.16~$\AA$
for the $4s$ and $4p$ states, and
1.22~$\AA$
for the $4d$ states,
anticipating that we will need to allow for 
$d$-states at higher pressures.
We perform the self-consistent calculation on our two-atom 
unit cell of arsenic using a 25$\times$25$\times$25
Monkhorst-Pack grid, which includes
the $\Gamma$ point.
A kinetic energy cutoff of 750~eV is used for the 
plane-wave expansion of the valence wavefunctions.  
A convergence threshold of $1.4\times10^{-8}$~eV
is set for the minimization of the total energy, and
a cold smearing~\cite{marzari_99} of 0.27~eV is used.
Once the ground state charge density of the system
has been determined,  using the {\tt PWSCF} code 
we then perform a non-self-consistent calculation 
in order to obtain a reference (\textit{ab initio}) band structure
against which the Wannier-interpolated band structure can be compared. 
To calculate the WFs, the self-consistent
potential is frozen and a non-self-consistent calculation
is performed in order to discretize the Bloch states over
a uniform mesh of 8$\times$8$\times$8 \mbox{$\q$-points}.
The strong localization of the MLWFs in real space, in addition to the generally-accepted
exponential fall-off of the WFs, ensures that 
a grid of this size is sufficient for our needs.~\cite{yates_07}
(In this work we adopt the convention used in Ref.~\onlinecite{yates_07}
of using the symbol $\q$ to denote grid points that
belong to the mesh upon which the Bloch states
have been stored.
This is in order to distinguish these specific
\mbox{$\q$-points} from
arbitrary sites, labeled
\mbox{$\k$-points},
at which the values are obtained
through interpolation.)
The Bloch states are thus stored on a relatively coarse
\textit{ab initio} grid. We have determined that we must use 9 WFs per
atom of the unit cell in order to ensure that we can capture
the character of the bands at higher pressures---this point
is elaborated upon in the supplementary material.~\cite{silas_12}
Thus we must construct
18 WFs and with this in mind, we calculate the first 30 bands at each
\mbox{$\q$-point}. The 18 WFs can then be constructed from these 30 bands
according to the disentanglement procedure prescribed by
Souza, Marzari and Vanderbilt.~\cite{souza_01}
This is performed using
the {\tt WANNIER90} code.~\cite{mostofi_08}
We obtain highly converged DOSs across the range of
pressures studied using a 400$\times$400$\times$400 interpolation (\mbox{$\k$-point}) mesh.
Although we use the \texttt{PWSCF} code in the LDA,
the configurations (the lattice parameters defining the structure of
arsenic at each pressure) used in this study were obtained from  
calculations using the Perdew-Burke-Ernzerhof generalized
gradient approximation~\cite{perdew_96} (abbreviated here as GGA-PBE)
for the exchange-correlation functional.~\cite{silas_yates_haynes_08}
We have confirmed that when using Wannier interpolation
to study the GGA-PBE configurations, it makes no
significant difference whether the ground state potential
has been set up using an LDA functional or a GGA functional.
%

\section{The Fermi Surface of Arsenic}
\label{sec:fs_of_As}

\begin{figure}[htpb]
\centering
\includegraphics[width=0.9\columnwidth]{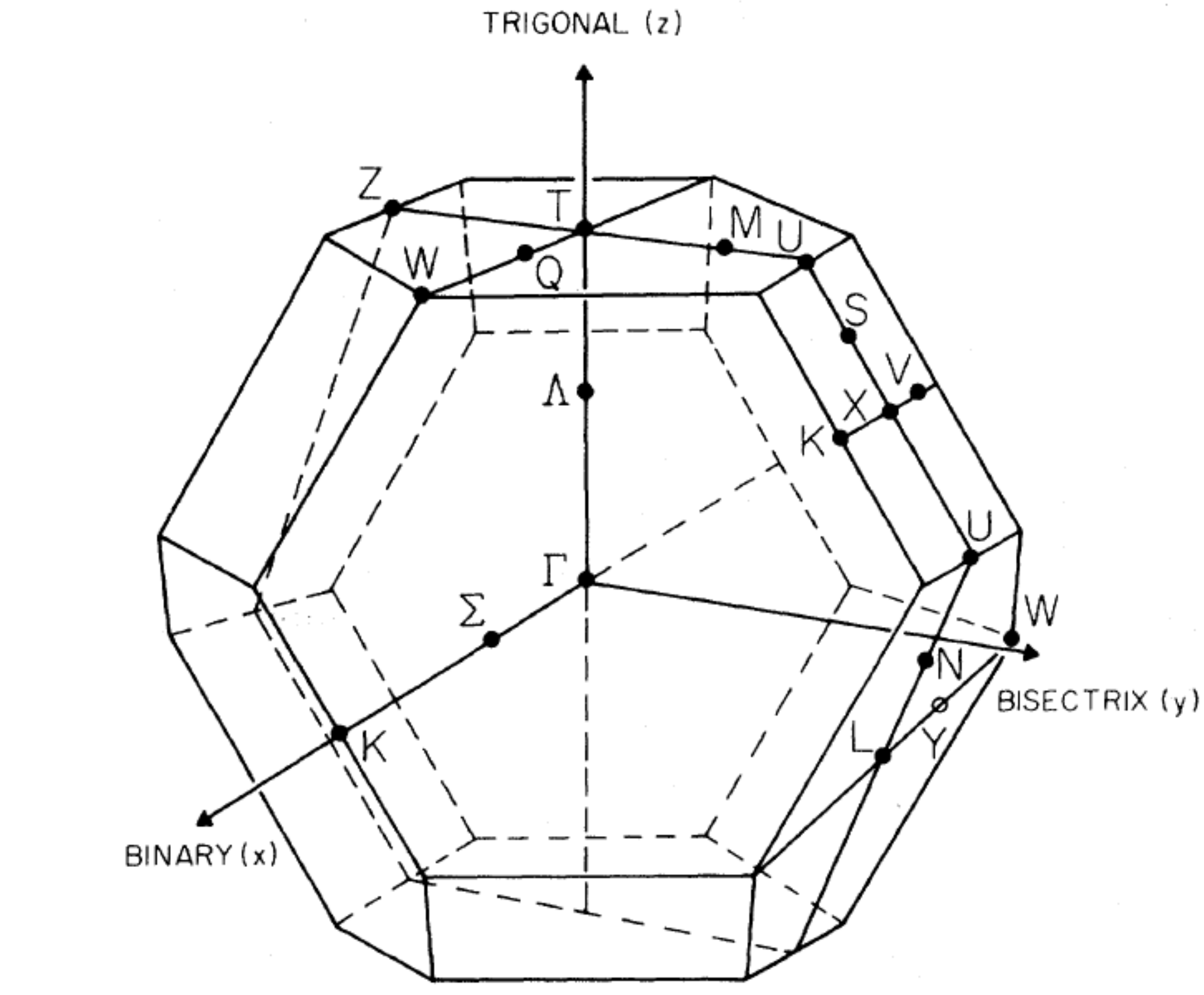}
\caption {\label{fig:a7bz1} The BZ of the primitive rhombohedral (A7) structure,
labeled with points and lines of symmetry~\cite{mhcohen_61,falicovandgolin_65}
and with the mutually orthogonal binary ($x$), bisectrix ($y$) and
trigonal ($z$) axes.  This figure is taken from Ref.~\onlinecite{cooper_71}.
The point along the TW line close to T is labeled as Q in this figure---this
point is usually called ``B'' in the literature and we refer to it throughout
this work as B. The special points appearing in the figure are defined explicitly
in the supplementary material.~\cite{silas_12}} 
\end{figure}

%
At 0~GPa, when it is in the A7 phase, arsenic
is a semimetal. Only the fifth and sixth
bands cross the Fermi level giving rise to the hole
and electron Fermi surfaces, respectively.
The BZ of the primitive
rhombohedral (A7) structure of arsenic is
presented in Fig.~\ref{fig:a7bz1}.
This figure was taken from
Ref.~\onlinecite{cooper_71}, though it was originally published by M. H. Cohen
in 1961,~\cite{mhcohen_61} and it is the same for arsenic as it is for
the other group--V semimetals, antimony and bismuth.
The special points of the rhombohedral BZ indicated
in Fig.~\ref{fig:a7bz1} have been written out explicitly
by Falicov and Golin,~\cite{falicovandgolin_65} and are given
in the supplementary material that accompanies this work.~\cite{silas_12}
%
%

\begin{figure}[htpb]
\centering
\includegraphics[width=0.8\columnwidth]{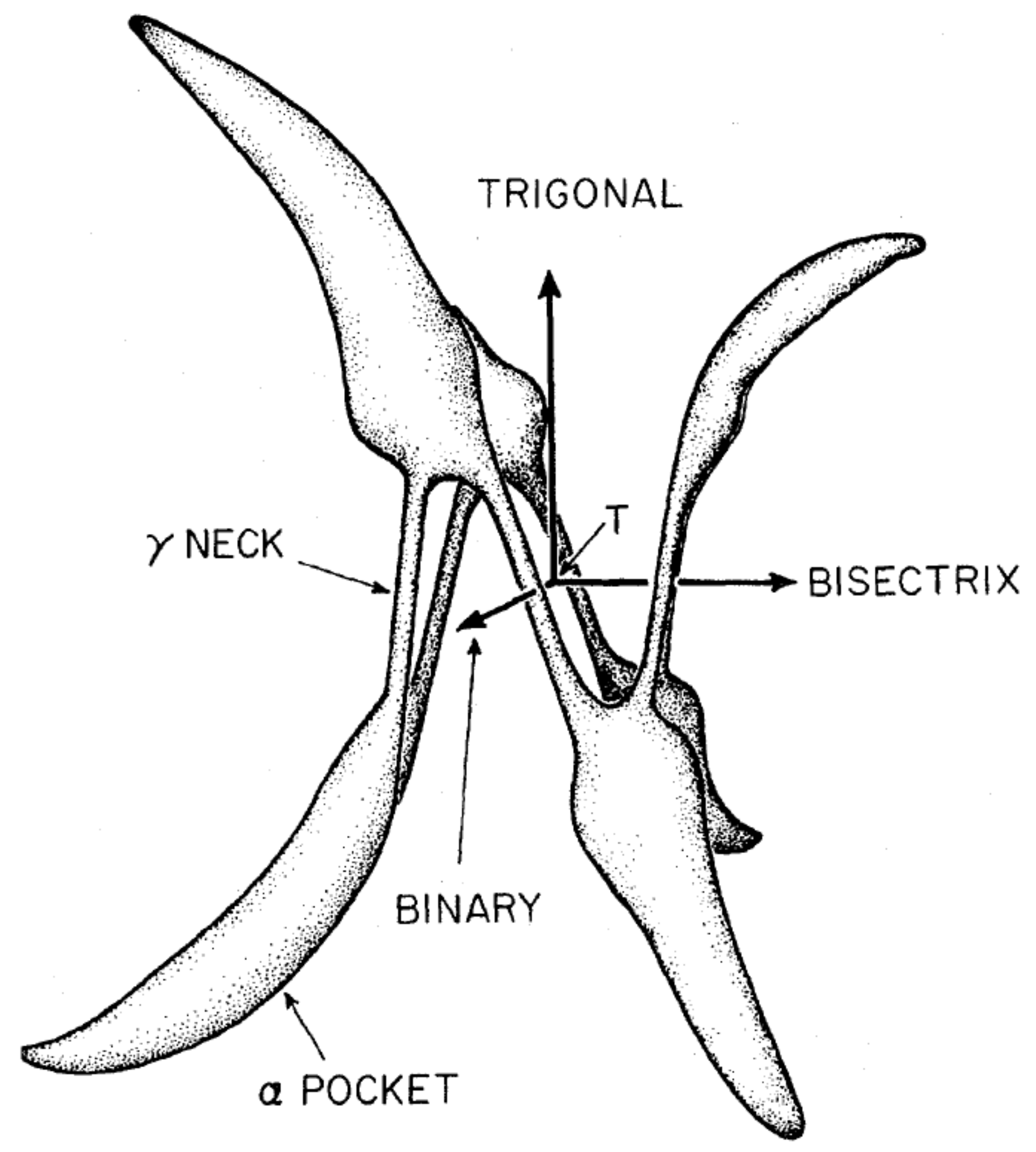}
\caption {\label{fig:hole_FS_CL} Lin and Falicov's~\cite{linandfalicov_66} hole crown,
taken from Ref.~\onlinecite{cooper_71}.  The hole Fermi surface
is centered at T. The mutually orthogonal binary~($x$), bisectrix~($y$) and
trigonal~($z$) axes are included.}
\end{figure}

\begin{figure}[htpb]
\vspace{-6mm}
\centering
\includegraphics[width=\columnwidth]{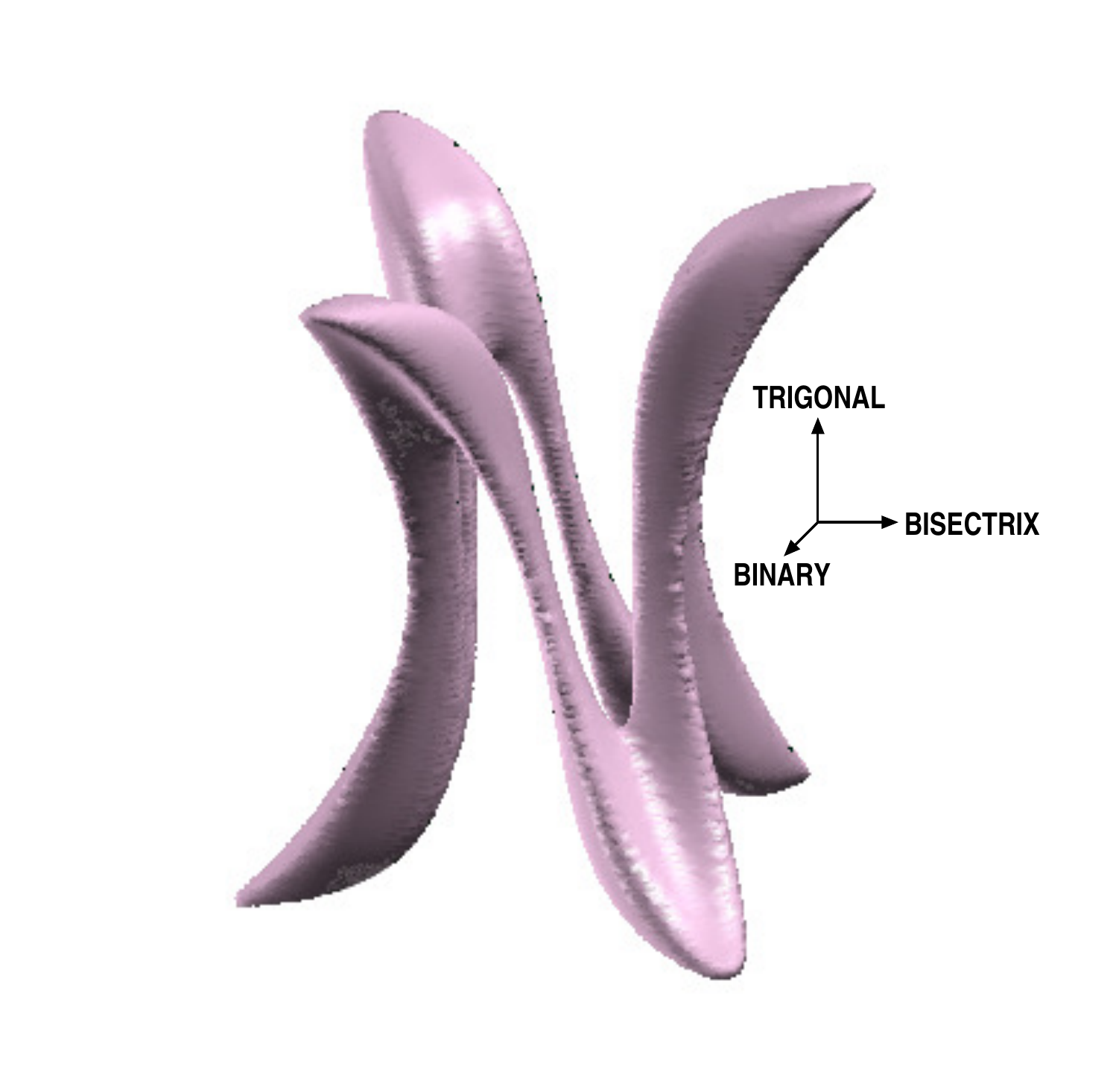}
\vspace{-12mm}
\caption {\label{fig:hole_FS_me} (Color online) The hole crown of arsenic at 0~GPa as it appears
in the reciprocal unit cell. The mutually orthogonal binary, 
bisectrix and trigonal axes are included. This has been
plotted using the \texttt{XCRYSDEN} package.~\cite{xcrysden}}
\vspace{-3mm}
\end{figure}

%
The hole Fermi surface, or hole ``crown'',
of arsenic depicted in Fig.~\ref{fig:hole_FS_CL}
was determined by Lin and Falicov via an empirically-based 
pseudopotential band structure calculation in 1966.~\cite{linandfalicov_66}
It is an artist's 
rendition, and it is still used today to illustrate the Fermi
surface of arsenic. 
We use Wannier interpolation to calculate the Fermi surface
of arsenic (at 0~GPa) for the first time since Lin and Falicov's
calculation.  From our 8$\times$8$\times$8 \textit{ab initio} grid,
we interpolate onto a 150$\times$150$\times$150 \mbox{$\k$-point}
grid spanning the BZ in an effort to capture the finest
details of the Fermi surface.  Given the results of this interpolation
and the Fermi energy computed in the initial self-consistent calculation,
the Fermi surface is plotted.
The hole Fermi surface of arsenic at 0~GPa resulting from our
Wannier interpolation is depicted in Fig.~\ref{fig:hole_FS_me}---this
is as it appears in the reciprocal unit cell.
This Fermi surface
does closely resemble Lin and Falicov's hole crown.
We will investigate how closely further below.
The hole Fermi surface, centered at the point T,
is composed of six lobe-like pockets
connected by six long thin cylinders
or necks, each of which is due to a point ``B''
of accidental degeneracy located along the TW line
and only slightly above the Fermi
level. (This accidental degeneracy is believed to be lifted
when the spin-orbit coupling is taken into account.~\cite{falicovandlin_66,
linandfalicov_66})
The lobes are each bisected by a plane containing
the trigonal ($\Gamma$T) and bisectrix ($\Gamma$U) axes
(by a mirror plane).  The maximum of the fifth band occurs
within the lobe and on the mirror plane at a point designated
as ``H''. 
\begin{figure}[htpb]
\centering
\includegraphics[width=\columnwidth]{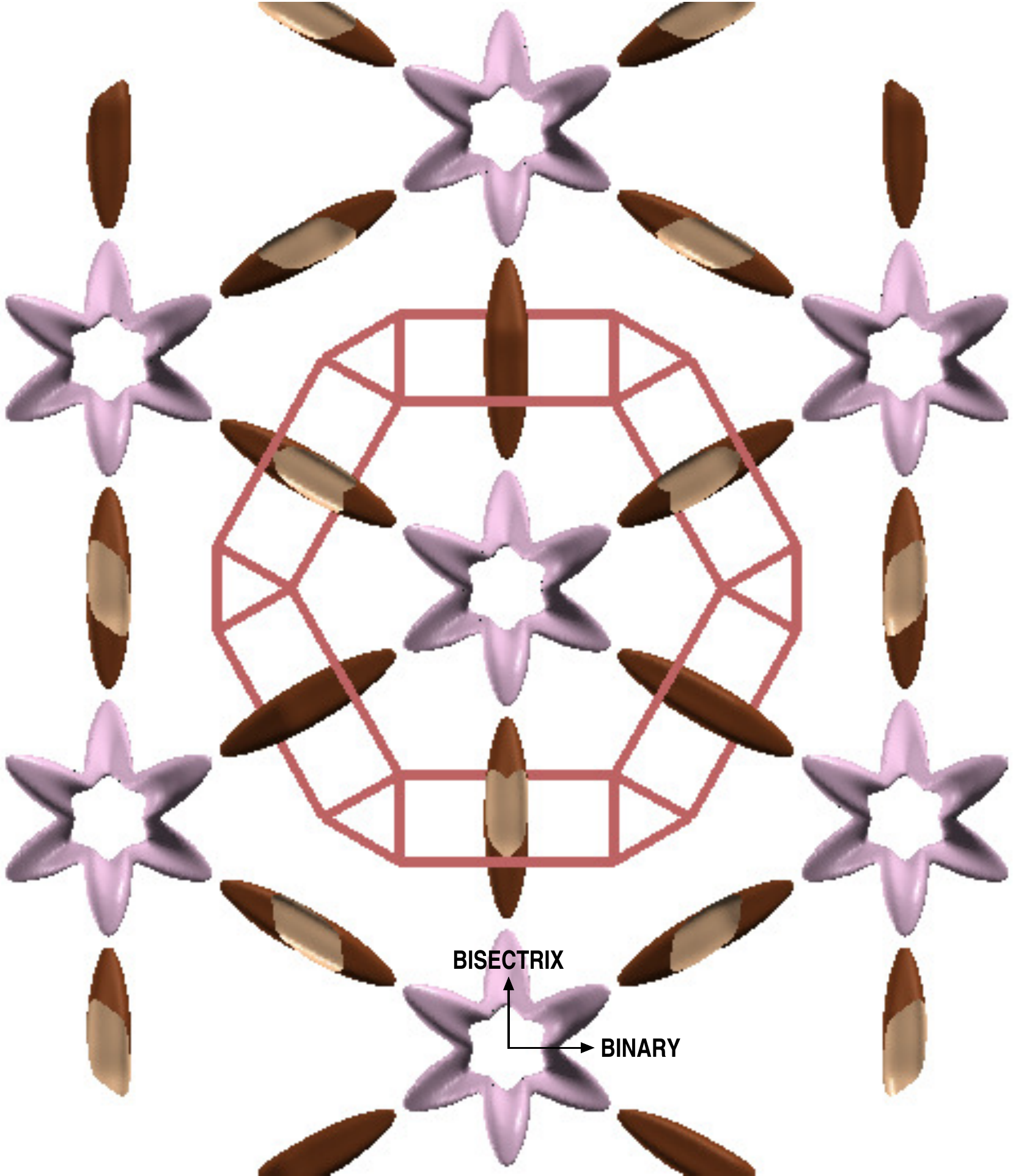}
\caption {\label{fig:multiple_hole_and_el} (Color online) The uncropped hole and electron
Fermi surfaces of arsenic at 0~GPa, where the first BZ is displayed.
The hole Fermi surface (in light gray/pink) is centered at the point T.  Three
of the six lobes shown are up, three down.
The electron Fermi surface (in dark gray/brown) is comprised of three
ellipsoid-like pockets centered at the three equivalent L points of the BZ.
The bisectrix axis (TU) is along the vertical, the 
binary axis (TW) along the horizontal.  The trigonal axis is through the origin and out 
of the page.}
\vspace{-2mm}
\end{figure}
The electron Fermi surface is
composed of three ellipsoid-like pockets centered at each of the
three equivalent L points of the BZ
(minimum of the sixth band occurring at L).
The merged hole and electron Fermi surfaces of arsenic at 0~GPa
are shown in Fig.~\ref{fig:multiple_hole_and_el},
where the binary (TW) axis lies along the horizontal and the bisectrix (TU)
axis lies along the vertical (see Figs.~\ref{fig:a7bz1} and~\ref{fig:hole_FS_CL}).
We will now go into more detail by looking at certain cross sections
of interest of the hole and electron Fermi surfaces of arsenic
at 0~GPa, and by comparing our results with those that are available
in the literature.  This will require interpolating onto two-dimensional
grids---slices through the BZ.
The Fermi contours resulting from the intersection of these
slices with the Fermi surface are subsequently plotted and examined.
We reveal next the intersection of one of these slices through the BZ
with the hole Fermi surface, resulting in the contour
displayed in Fig.~\ref{fig:hole_3}.  Here we have sliced
the hole Fermi surface parallel to the trigonal-bisectrix plane and passing
through B.  We include as an inset of this figure the equivalent contour
obtained by Lin and Falicov in 1966.~\cite{linandfalicov_66}  Our Fermi contour is considerably more
detailed than that of Lin and Falicov.
The contours resulting from the other slices we inspected are presented
in the supplementary material,~\cite{silas_12} 
but all results are summarized
in Table~\ref{table:areas} and contrasted against any data previously
published in the literature.
\begin{figure}[htpb]
\centering
\includegraphics[width=\columnwidth]{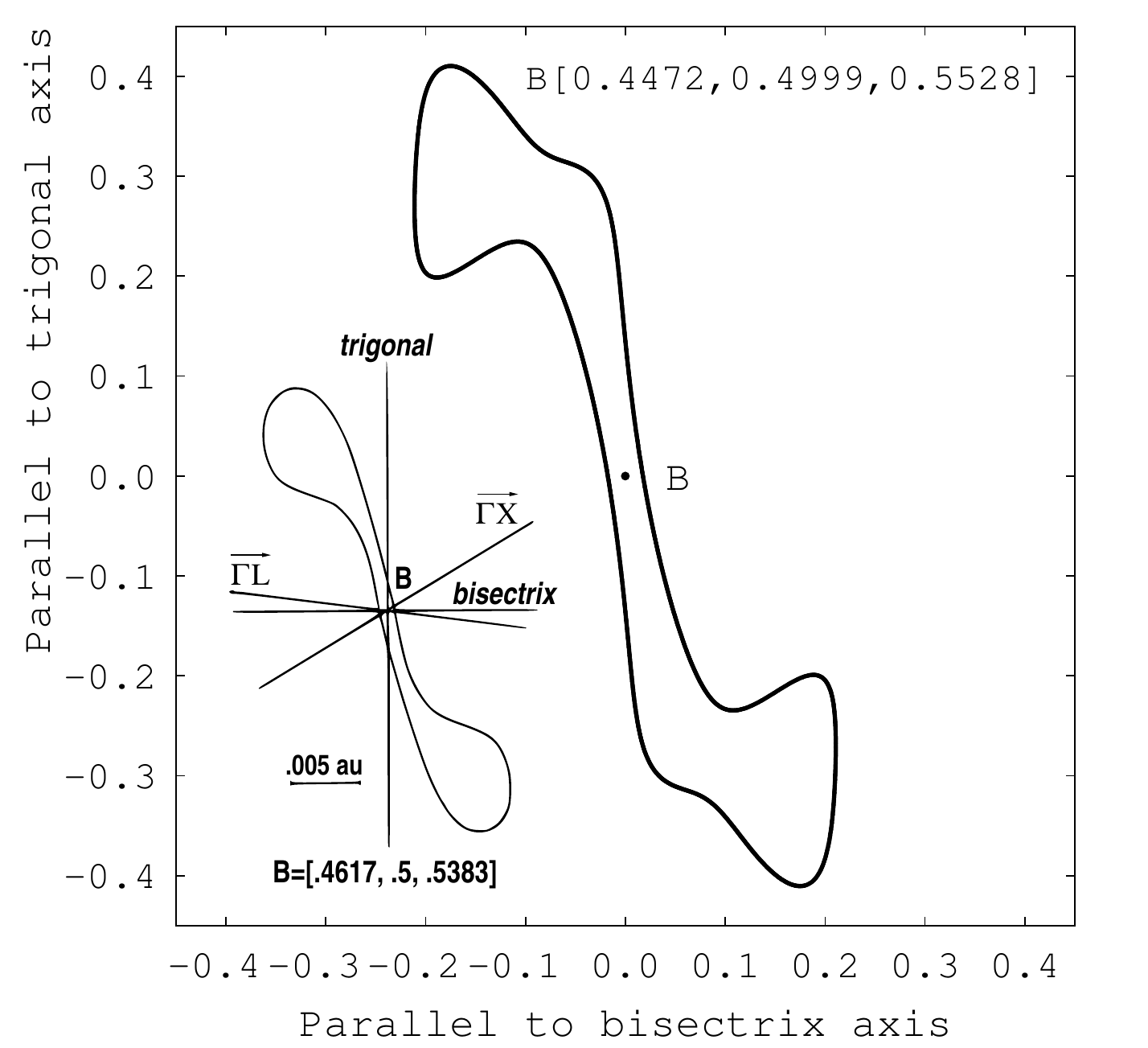}
\caption {\label{fig:hole_3} Hole pocket: trigonal-bisectrix plane through B. 
This figure has been obtained
using a Fermi energy recomputed from the Wannier-interpolated
DOS of A7 arsenic at 0~GPa, presented in Sec.~\ref{sec:DOS}.
All distances are in $\AA^{-1}$.
The
inset is the analogous cross section as calculated by Lin and Falicov.~\cite{linandfalicov_66}}
\end{figure}
%

\begin{table*}[htpb]
\begin{minipage}{\textwidth}
\centering
\caption{Features of the electron and hole Fermi surfaces of arsenic.}
\label{table:areas}
\tiny{
\begin{tabular}{lcccccccc}
\hline
\hline
Electron Fermi surface.  All areas are in $\AA^{-2}$\TM & & Ref.~\onlinecite{linandfalicov_66} & &
Ref.~\onlinecite{priestley_67} & & Ref.~\onlinecite{cooper_71} & & This work\footnote{\tiny {
These values have been obtained using a Fermi energy recomputed
from the Wannier-interpolated DOS of A7 arsenic at 0~GPa,
presented in Sec.~\ref{sec:DOS}.}}\\
\B  & & (Theory) & & (Experiment) & & (Experiment) & & \\
\hline
Area normal to the binary [see Fig.~\ref{fig:electron_1}\footnote{\tiny{Figures for which the numbering begins with an ``S'' are found in the supplementary material provided.}}]: trigonal-bisectrix plane through L\T & & 0.0571\footnote{\tiny{Electron pockets: Fermi energy
 fixed so as to fit the minimum area for magnetic fields in the trigonal-bisectrix plane.}} & & 0.0732 $\pm$ 0.0002 & & 0.0733 & & 0.088\\
\hspace{0.44 in} Tilt angle\footnote{\tiny{Angle convention followed throughout is that of Ref.~\onlinecite{cooper_71} (see also
 Refs.~\onlinecite{falicovandlin_66,windmiller_66,priestley_67,ih_70}):
angles are measured in the trigonal-bisectrix
plane or some parallel plane and with respect to the vertical---the trigonal axis ($\Gamma$T) or some parallel line. Positive
rotations are in the sense from $\Gamma$T to $\Gamma$X in
the first quadrant of the coordinate
system.}} & & $-8^\circ$ & & $-9.0 \pm 0.2^\circ$ & & & & $-10.6^\circ$\\
Area normal to the trigonal [see Fig.~\ref{fig:electron_2}]: binary-bisectrix plane through L & & 0.0643 & & 0.0746 $\pm$ 0.0002 & & 0.0746 & & 0.096\\
Area normal to the bisectrix [see Fig.~\ref{fig:electron_3}]: trigonal-binary plane through L\B & & & & & & 0.0202 & & 0.026\\

\hline
\hline

Hole Fermi surface.  All areas are in $\AA^{-2}$\TM\B & & & & & & & &\\

\hline

Area normal to the binary [see Fig.~\ref{fig:hole_1}]: trigonal-bisectrix plane through $\Gamma$\T & & 0.0343 & & 0.01422 $\pm$ 0.00002 & & 0.0275 & & 0.027\\
\hspace{0.44 in} Tilt angle & & $\sim +44^\circ$ & & $+37.25 \pm 0.1^\circ$ & & $+37.3 \pm 1.5^\circ$ & & $+37.8^\circ$\\
Area normal to the trigonal [see Fig.~\ref{fig:hole_2}]: binary-bisectrix plane through T & & $(2.464 \times 10^\rm{-4})$\footnote{\tiny{Hole pockets: Fermi energy fixed so as to fit this area.}} & & $(2.453 \pm  0.007) \times 10^{-4}$ & & 0.0179 & & $2.1 \times 10^{-3}$\\
\hspace{0.44 in} Tilt of necks [see Fig.~\ref{fig:hole_3}] & & $-11^\circ$ & & $-9.6 \pm 0.1^\circ$ & & & & $\sim -9.7^\circ$\\
Combined area normal to the bisectrix [see Fig.~\ref{fig:hole_4}]: trig.-bin. plane through T\B & & & & & & 0.0367 & & 0.030\\

\hline
\hline

\end{tabular}
\begin{tabular}{lcccc}
Special \mbox{$\k$-points} H and B. Coordinates are fractional w.r.t. reciprocal lattice vectors.\TM\B & & Ref.~\onlinecite{linandfalicov_66} & & This work\\
\hline
1 of 6 equivalent H points---maximum of the $5^\rm{th}$ band (occurs in mirror plane)\T & & [0.2043, 0.3758, 0.2043] & & [0.2050, 0.3753, 0.2050]\\
1 of 6 equivalent B points---point of accidental degeneracy along TW line\B & & [0.4617, 0.5, 0.5383] & & [0.4472, 0.4999, 0.5528]\\

\hline
\hline
\end{tabular}
}
\end{minipage}
\end{table*}

\section[Pressure Dependence of the Fermi Surface of Arsenic]
{Pressure Dependence of the Fermi Surface of Arsenic: 
the A7~$\to$~SC Phase Transition}
\label{sec:pressure_dependence_of_fs_of_As}

%
\begin{figure*}[htpb]
\centering
\includegraphics[width=0.93\textwidth]{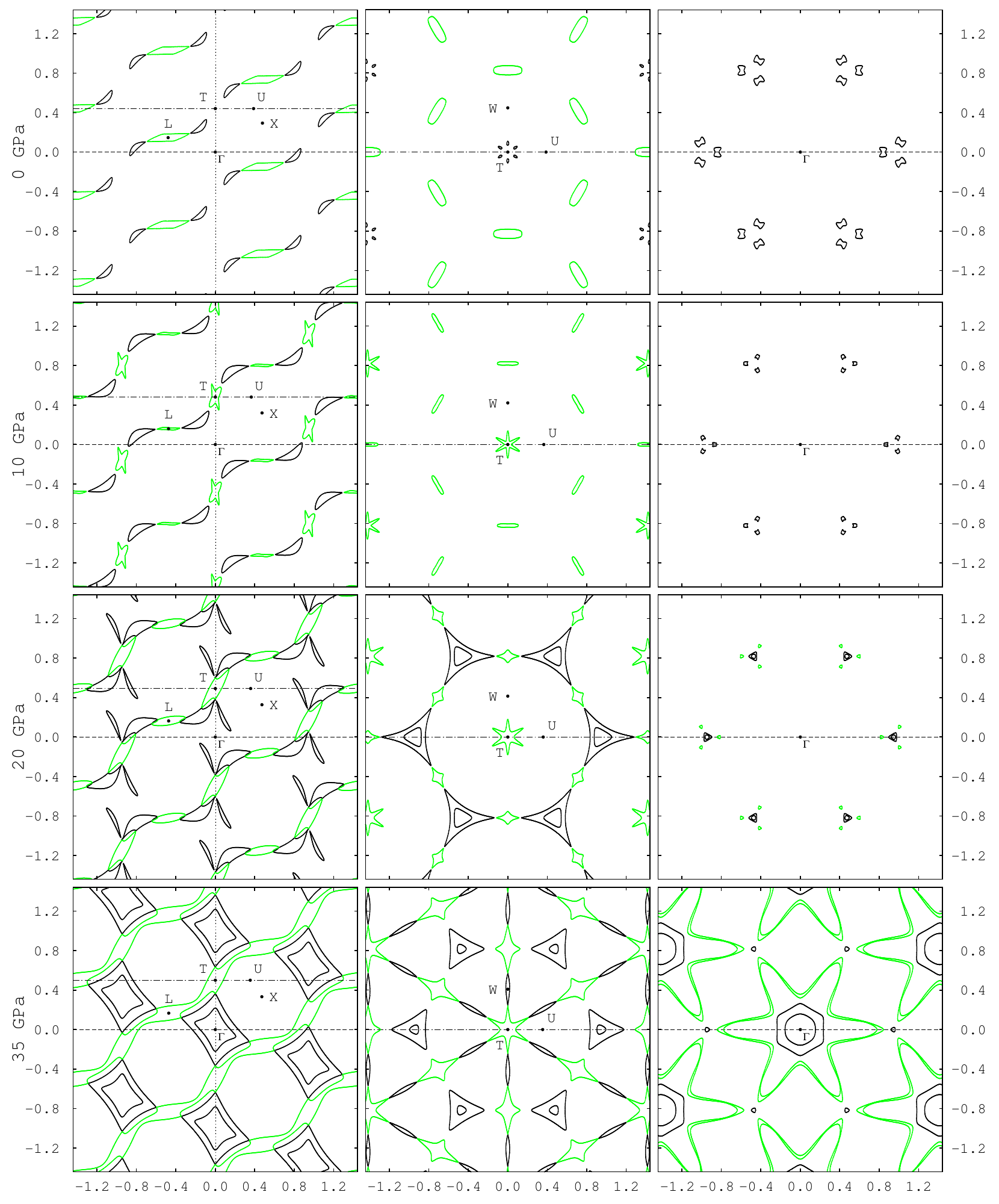}
\caption {\label{fig:0-35GPa} (Color online) Pressure dependence of the Fermi surface of arsenic: \mbox{0--35GPa}.
For the left-hand column, the BZ has been intersected by the trigonal-bisectrix plane;
the middle column by the binary-bisectrix plane through T; the right-hand column by
the binary-bisectrix plane through $\Gamma$.  The bisectrix axis (the dotted-dashed line)
is along the horizontal
for all three columns.  The trigonal axis (the finely dotted line)
is along the vertical for the left-hand column.
The binary axis is along the vertical for the middle
and right-hand columns.  All distances are
fractional with respect to the length of the reciprocal lattice
vector---this is to enable comparisons across pressures. Hole contours are in black, electron contours in gray (green).
The cross section of
the hole crown centered at T can be seen in the top left panel.
The A7~$\to$~sc phase transition takes place at a pressure between
those represented in the bottom two rows of the figure.
In the bottom row (at 35~GPa) arsenic is simple cubic---the use of a two-atom
unit cell causes the folding evidenced especially in the bottom left and bottom right panels.}
\end{figure*}
We illustrate in Fig.~\ref{fig:0-35GPa} the evolution of the electron and hole
Fermi surfaces\footnote{Thus far,
our Fermi energies have been obtained using the
PWSCF code.~\cite{pwscf}  Unless otherwise indicated, our depictions
of the Fermi surface of arsenic have also been obtained
using these values.  Our measurements of the cross sections
of the Fermi surface of arsenic at 0~GPa however have been
calculated using a more precise value of the Fermi energy,
recomputed from a Wannier-interpolated
DOS we present in Sec.~\ref{sec:DOS}.}
of arsenic as it undergoes the A7~$\to$~sc
phase transition.  This figure is complemented by the corresponding
Animations~01--03,\cite{animation1,animation2,animation3}
of the supplementary material provided.~\cite{silas_12}
These animations show the phase transition happening over a finer resolution
of pressures than is shown in Fig.~\ref{fig:0-35GPa},
and thus give a better idea of what happens to the Fermi
surface through the~transition.
Fig.~\ref{fig:0-35GPa} consists of three columns, each representing
a different cross section of the BZ. The four rows of the 
figure refer to four different pressures: 0~GPa, 10~GPa, 20~GPa and 35~GPa.
In all three columns, the bisectrix axis lies along the horizontal.
For the left-hand column the trigonal axis lies along the vertical, whereas
for the other two columns the binary axis lies along the vertical.
Arsenic is in the A7 phase for the first three rows---it is in the 
sc phase in the fourth row. Hole
surface contours are in black and electron surface contours
are in gray (green).
This investigation complements our earlier studies of the A7~$\to$~sc
phase transition of arsenic.~\cite{silas_yates_haynes_08}
The configurations resulting from geometry optimizations performed 
on arsenic subjected to pressures up to 200~GPa in that work act as
the starting point here for our studies of arsenic using Wannier interpolation.
We present our findings concerning the higher-pressure
phase transitions of arsenic in the supplementary
material provided.~\cite{silas_12}
We expect the A7~$\to$~sc phase transition to occur at
$28\pm1$~GPa.
We first discuss the left-hand column of Fig.~\ref{fig:0-35GPa}, in which
the A7~$\to$~sc phase transition is most
noticeable.  The left-hand column of Fig.~\ref{fig:0-35GPa} corresponds
to the intersection of the BZ with the trigonal-bisectrix
plane through $\Gamma$ (it is a mirror plane and contains the points
$\Gamma$, T, L, X and U as indicated in the figure).
The trigonal axis is the finely dotted vertical line.  The other
two lines in the diagrams on the left serve to compare orientations with
the other two columns.
As we scan from top to bottom in this first column
 (corresponding to Animation~01\cite{animation1}), we see that the hole
surface initially centered on T distorts in such a way that by the 
bottom panel there is a folded cubic surface (cut along the body diagonal)
centered at $\Gamma$.  Arsenic at 35~GPa is in the sc phase.
The surface is folded due to the fact that we have two atoms in the unit cell.
We notice too that by 10~GPa an electron pocket has
opened up at T (it has actually already opened up by 8~GPa).
Our Fermi surfaces
fold from about 27~GPa, agreeing with our expected transition pressure of $28\pm1$~GPa.
For the middle column of Fig.~\ref{fig:0-35GPa}
 (corresponding to Animation~02\cite{animation2}), the BZ has
been sliced by the binary-bisectrix plane through T. This slice thus also
contains points W and U as indicated.
The dotted-dashed line in these figures is the bisectrix axis---as a visual aid for
orientation purposes, the bisectrix axis can be compared for corresponding
panels of the left-hand and middle columns.
The phase transition, again from about 27~GPa, seems to happen
when the hole pockets that have opened up at W are just touching the
electron pockets appearing on either side of them.
In the right-hand column of Fig.~\ref{fig:0-35GPa}
 (corresponding to Animation~03\cite{animation3}), the BZ
has been sliced by the binary-bisectrix plane through $\Gamma$.  The
dashed lines appearing in the panels of this column correspond to the
dashed lines appearing in the panels of the left-hand column of the figure.
At about 27~GPa, we once again see the folding of the individual electron
and hole Fermi surfaces indicating that arsenic is now in the sc
phase.  
We have mentioned that the folding of the Fermi surfaces
seen in the bottom row of Fig.~\ref{fig:0-35GPa} is due to 
the fact that we are using the two-atom unit cell to model arsenic
over the entire range
of pressures studied, whereas in the sc phase the primitive
cell contains one atom. The \textit{unfolded} Fermi surfaces can be inspected 
by comparing the bottom row of Fig.~\ref{fig:0-35GPa}
with Fig.~S9 of the supplementary material,~\cite{silas_12}
in which are presented the corresponding results
for sc arsenic at 35~GPa using a one-atom
primitive cell.
The BZ
for the one-atom primitive sc structure
can be found for example in Ref.~\onlinecite{mhcohen_64}
or more recently in Ref.~\onlinecite{martin}.
Determining the pressure at which the A7~$\to$~sc transformation of arsenic happens
is difficult due to the nature of this semimetal to metal phase transition,
which in our earlier study we found to be of second order.~\cite{silas_yates_haynes_08} 
In that work, we investigated the pressure dependence of the lattice parameters of
arsenic, and we determined the transition pressure according to when the internal
coordinates had reached their high-symmetry values.  However as the internal coordinates
reached their high-symmetry values at two different pressures, there was still
some question as to when the transition actually happens.  Using Wannier interpolation
we have observed the evolution of certain cross sections of interest of the
Fermi surfaces across the transition. Wannier interpolation allows us to say
that the folding of the Fermi surfaces is perhaps
the best indication of when the A7~$\to$~sc
phase transition has occurred.
%

\section{Wannier-Interpolated Densities of States 
\mbox{of A7 and SC Arsenic}}
\label{sec:DOS}

\begin{figure}[htpb]
\centering
\includegraphics[width=\columnwidth]{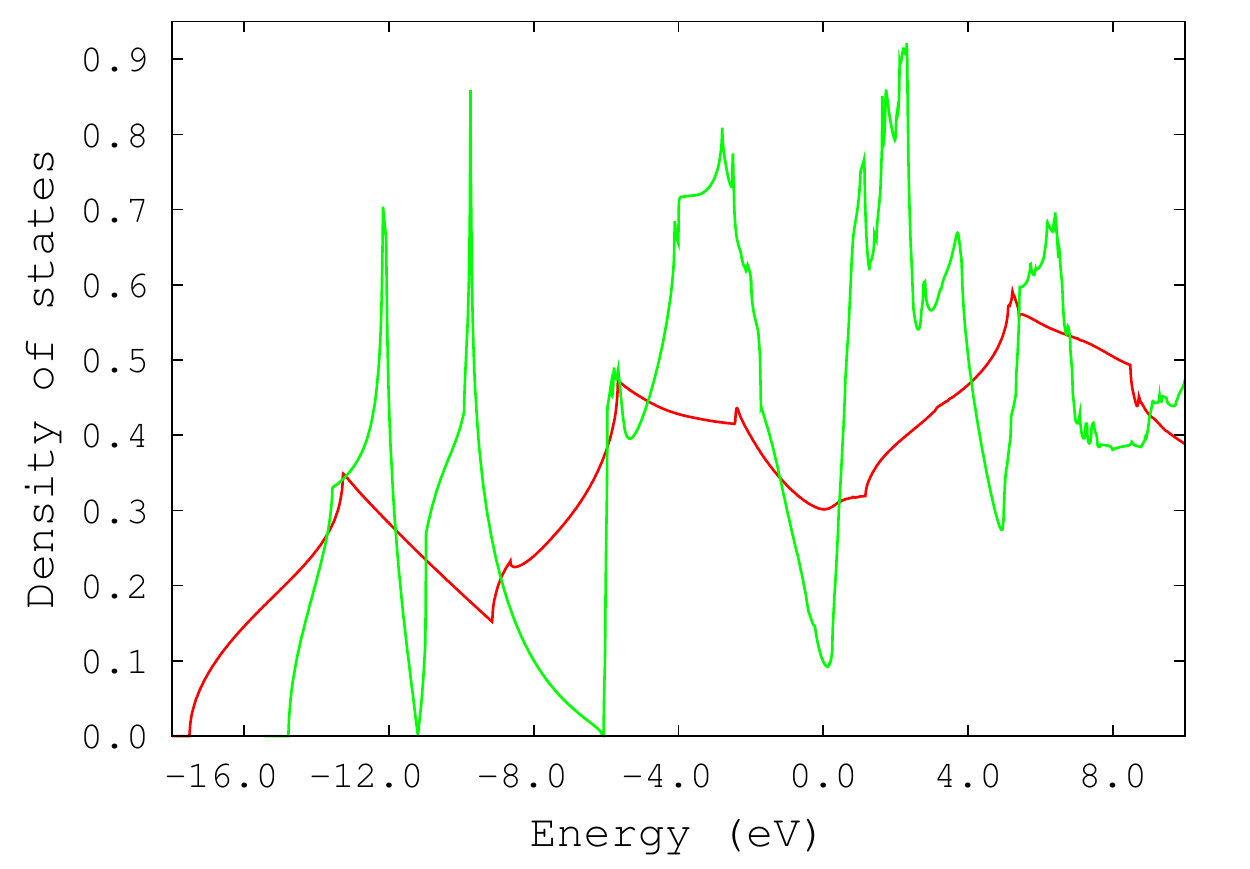}
\caption {\label{fig:DOS-0-and-35GPa} (Color online) Wannier interpolation combined with
adaptive smearing~\cite{yates_07} using cold smearing:~\cite{marzari_99}
comparing the DOSs of A7 and sc
arsenic---the pressures are 0~GPa (gray/green) and 35~GPa (black/red), respectively.
The DOSs are referenced to the Fermi level in both cases.
This figure can be compared to that which appears in Ref.~\onlinecite{mattheiss_86}.
Our results agree qualitatively with those.  It is evident that
Wannier interpolation combined with adaptive smearing enables features of the DOS to
be captured on the very finest of scales.
The DOS at the Fermi level can be seen to increase appreciably through the A7$~\to~$sc
transition, as arsenic changes from a semimetal to a metal.
}
\end{figure}

%
Wannier-interpolated DOSs for arsenic in the A7 phase at 0~GPa
and in the sc phase at 35~GPa obtained using
the {\tt WANNIER90} code~\cite{mostofi_08}
are presented in Fig.~\ref{fig:DOS-0-and-35GPa}.
This figure can be compared with that which appears
in Ref.~\onlinecite{mattheiss_86}.
Our results agree at least qualitatively with those
(the pressures corresponding to the DOSs in that
publication are not specified), the only other
DOS calculations of arsenic of which we are aware.
It is evident that
Wannier interpolation combined with adaptive smearing enables highly-accurate
DOSs, allowing even the very finest of features to be captured.
As we would expect, there is a dramatic increase in the DOS
at the Fermi level as the pressure is increased through the 
A7~$\to$~sc phase transition, during which arsenic changes from a semimetal
to a metal.
We have used our DOS calculations to recompute the Fermi energy
of A7 arsenic at 0~GPa.  We obtain a value
that is 0.01~eV higher
than that resulting from our self-consistent calculation using the
PWSCF code, moving the Fermi level toward the minimum
in the DOS that appears nearby.
\begin{figure}[htpb]
\centering
\includegraphics[width=\columnwidth]{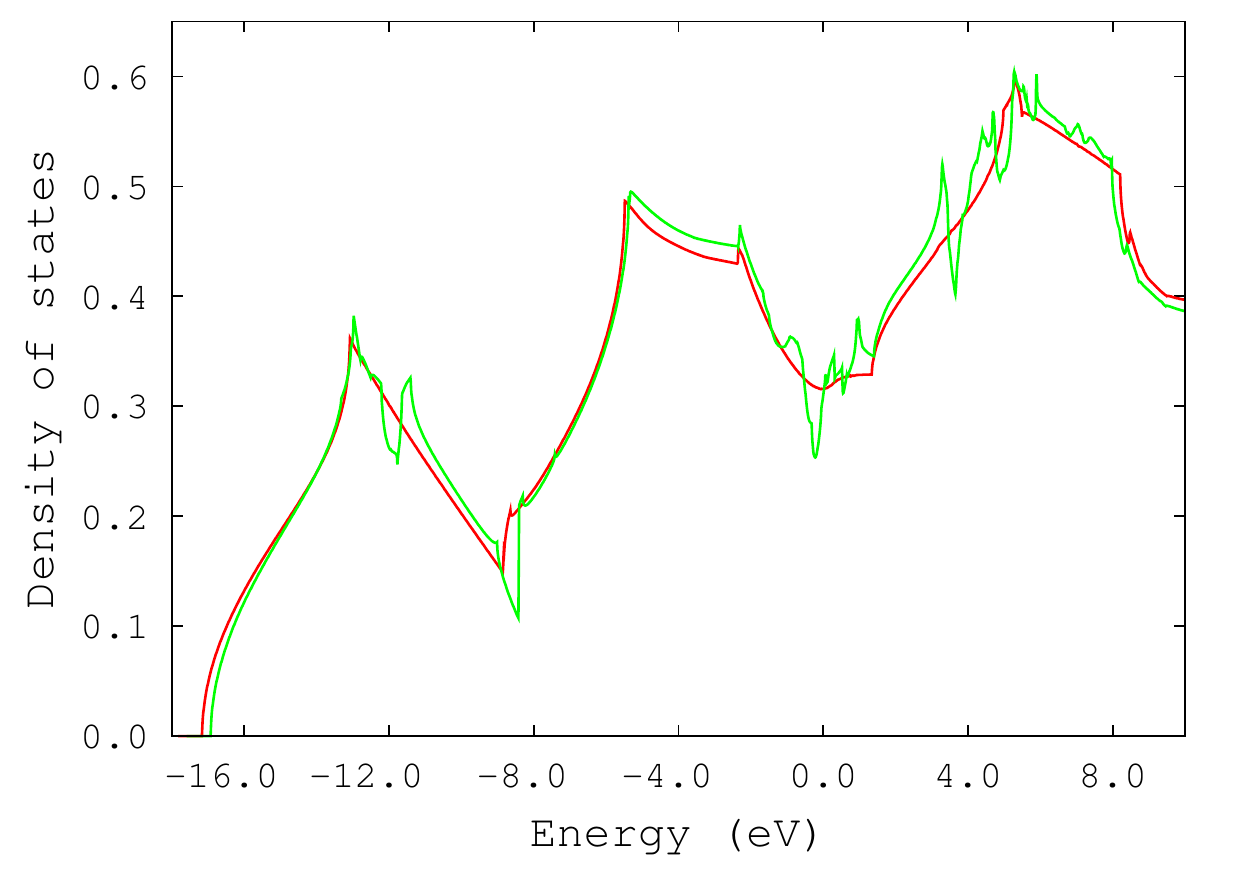}
\caption {\label{fig:DOS-25+29GPa} (Color online) Evolution of the DOS of arsenic
across the sc~$\to$~A7 phase transition.
The DOSs of arsenic
at 29~GPa (black/red) and at 25~GPa (gray/green) are referenced to the Fermi level and
superimposed. It can be seen that as the pressure is lowered, 
the onset of the Peierls-type cubic to rhombohedral distortion is
signified by the emergence of van Hove singularities in the DOS.
}
\end{figure}
Next we inspect the evolution of the DOS
of arsenic in the immediate
vicinity of the A7~$\to$~sc transition, and present our results
in Fig.~\ref{fig:DOS-25+29GPa}.
In this figure, the DOSs for arsenic at 29~GPa
and 25~GPa are referenced to the Fermi level and superimposed.
Changes in the DOS are observed more clearly as the pressure
is lowered such that arsenic transitions from the higher-symmetry sc phase
to the lower-symmetry A7 phase (the sc~$\to$~A7 transition). 
Thus we see from this figure that the onset of
the Peierls-type cubic to rhombohedral distortion
is signified by the emergence of van Hove singularities in the DOS,
especially around the Fermi level.
The rapidly changing DOS at the Fermi level explains why such high levels of convergence
are required when studying this transition.~\cite{silas_yates_haynes_08}
In our earlier studies of arsenic,~\cite{silas_yates_haynes_08}
 we had found that the rhombohedral angle $\alpha$ reaches
its high-symmetry value at a lower pressure than does
the atomic positional parameter $z$, and that 
the electronic change occurring
 as the pressure is increased through the A7~$\to$~sc
transition appeared to be driving the atomic positional parameter $z$
to its high-symmetry value.
A close inspection of the evolving DOS however, which can be seen
in the supplementary material, reveals that the electronic change
is instead driving $\alpha$ to its high-symmetry value.
%

\section{Conclusions}
\label{sec:conclusion}

%
Wannier interpolation has enabled us to calculate the Fermi
surface of arsenic at ambient pressures and through
the A7~$\to$~sc phase transition, obtaining
first principles accuracy at vastly reduced computational cost.
We found that the folding of the Fermi surfaces
is perhaps the best indication of when the 
A7~$\to$~sc phase transition has occurred.
We further determined from our studies of the Wannier-interpolated DOS 
of arsenic over the range of pressures in the vicinity of the A7~$\to$~sc
phase transition, that the onset of the Peierls-type cubic to rhombohedral distortion
is signified by the appearance of emerging van Hove singularities in the DOS especially
around the Fermi level.
The rapidly changing DOS at the Fermi level explains why such high levels of convergence
are required when studying this transition, as we had found in our earlier studies of arsenic.\cite{silas_yates_haynes_08} 
Arsenic provides us with an example of a challenging system that
is ideally suited to demonstrating the power of the Wannier interpolation 
technique.  This methodology furthermore enables a novel approach to
the study of semimetal to metal phase transitions, such as the A7~$\to$~sc
transition of arsenic, or in fact of any phase transitions involving a metal. 
The technique is a particularly valuable one for the
accurate determination of Fermi surfaces, as well as for the calculation of
highly resolved DOSs.  
%

\section{Acknowledgments}

We thank Richard Needs for helpful discussions.  Computing resources
were provided by the University of Cambridge High Performance Computing Service (HPCS).
P.D.H. and J.R.Y. acknowledge the support of Royal Society
University Research Fellowships.

\appendix*

\section{Evolution of the Fermi surface of arsenic through the rhombohedral to
simple-cubic phase transition: a Wannier interpolation study---supplementary information}
\label{SI}

\setcounter{figure}{0}
\makeatletter
\renewcommand{\thefigure}{S\@arabic\c@figure} 

\setcounter{section}{0}
\makeatletter
\renewcommand{\thesection}{S\@Roman\c@section} 

We review the Brillouin zone (BZ) corresponding
to the two-atom primitive 
cell of rhombohedral (A7) arsenic, 
defining explicitly the 
special $\k$-points involved.  
This is followed by a discussion
concerning the choice of the number of Wannier functions
to construct in order to set up an ``exact tight-binding''
model at each given pressure from which 
Wannier interpolations can be performed.
We elaborate on the features of the electron and hole Fermi surfaces
of A7 arsenic at 0~GPa.
We investigate the case of simple-cubic (sc) arsenic
at 35~GPa using the one-atom primitive cell in order
to inspect the ``unfolded'' Fermi surfaces, and
to compare with the folded ones resulting from the
use of the two-atom unit cell.
We present the results of our studies
of arsenic at pressures beyond which it
is no longer in the sc phase,
where once again the two-atom unit cell is employed.
Finally, we elaborate on our studies of the density of states (DOS)
of arsenic as it undergoes the A7~$\to$~sc
 (space groups \textit{R$\bar{3}$m}~$\to$~\textit{Pm$\bar{3}$m})
 phase transition,
and present DOSs for A7, sc and body-centered-cubic (bcc) arsenic
(space group \textit{Im$\bar{3}$m}).
These DOSs are contrasted against the band structures from 
which they originate. 
%

\section{The Rhombohedral Brillouin Zone}
\label{sec:A7bz}

\begin{figure}[htpb]
\centering
\includegraphics[width=0.9\columnwidth]{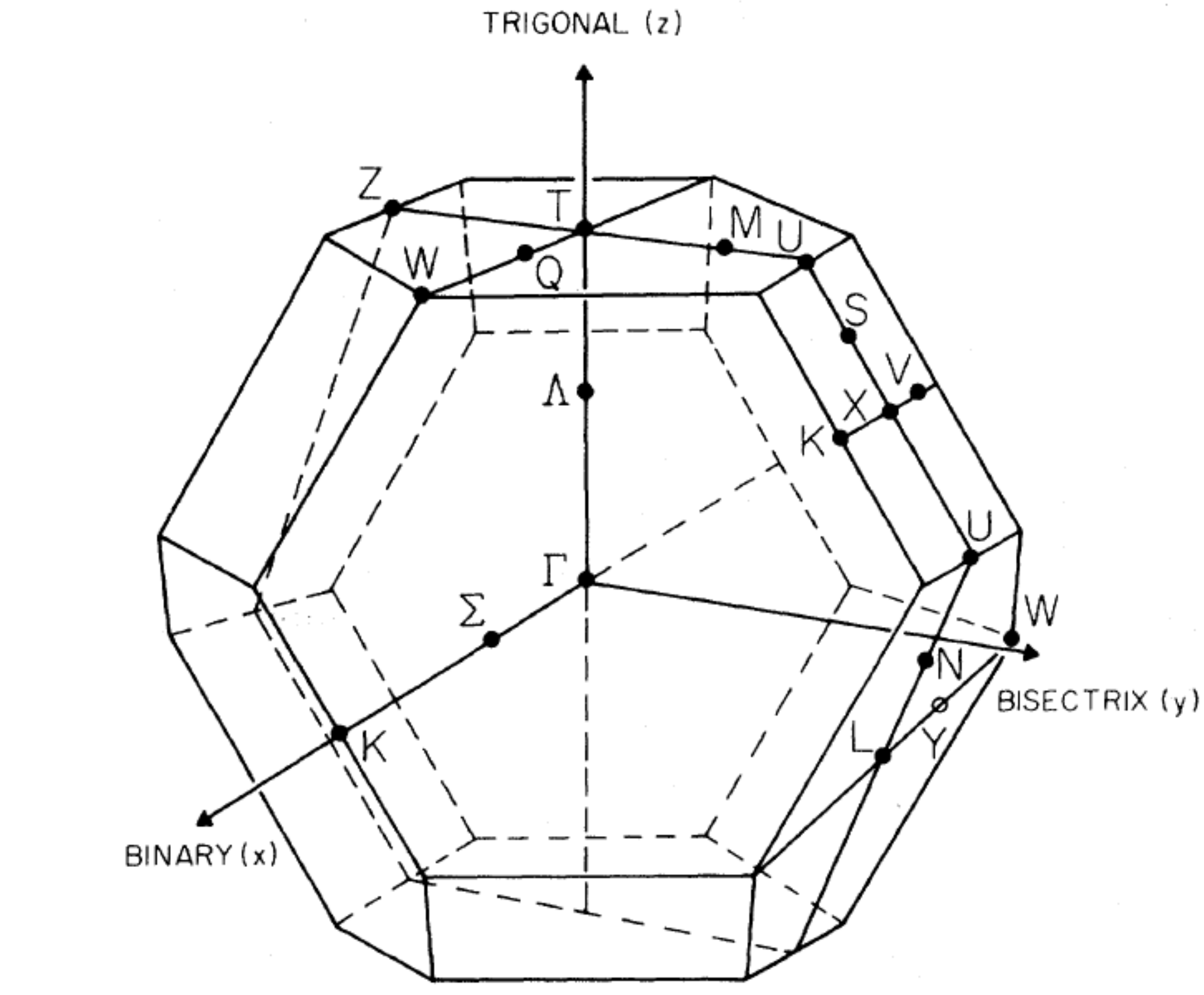}
\caption {\label{fig:a7bz2} \footnotesize The BZ of the primitive rhombohedral (A7) structure,
labeled with points and lines of symmetry~\cite{mhcohen_61,falicovandgolin_65}
and with the mutually orthogonal binary ($x$), bisectrix ($y$) and 
trigonal ($z$) axes.  This figure is taken from Ref.~\onlinecite{cooper_71}.}
\end{figure}

%
The real-space primitive cell of A7 arsenic contains two atoms and is described
by three lattice parameters: the length of the lattice vectors, $a$, the
angle $\alpha$ between each pair of lattice vectors, and the atomic positional
parameter $z$, where $z < 1/4$, which determines the fractional positions of the two atoms
along the cell's body diagonal. 
The BZ of the primitive rhombohedral structure---it is the same for
each of the group--V semimetals: arsenic, antimony and
bismuth---is presented
in Fig.~\ref{fig:a7bz2}.~\footnote{References
in which the numbering is not preceeded by an ``S'' refer to 
Figures, Sections or Tables that appear in the accompanying 
paper (Ref.~\onlinecite{silas_haynes_yates_12}).}
This figure was taken from
Ref.~\onlinecite{cooper_71}, though it was originally published by M. H. Cohen
in 1961.~\cite{mhcohen_61} 
Cohen, Falicov and Golin~\cite{mhcohen_64} showed that much of the band
structure of the group--V semimetals can be explained directly from
their crystal structure.
The parameter $\epsilon$ as defined in Ref.~\onlinecite{falicovandgolin_65}
describes the shear that distorts the A7 structure
from the sc structure. It is related to the rhombohedral
angle $\alpha$ thus:
\vspace{2mm}
\begin{equation}
 \label{eq:epsilon}
   \epsilon = \frac {\left[ 1 - \left( 1 + \cos \alpha
         - 2 \cos ^2 \alpha \right) ^{1/2} \right]} {\cos \alpha}.
\vspace{2.5mm}
\end{equation}
The structure of A7 arsenic can be pictured by imagining
the interpenetration of two face-centered-cubic lattices (each one
contributing an atom to the two-atom A7 primitive cell) offset from
each other such that $z < 1/4$ and to which
a shear has been applied in the [111] direction (along the trigonal axis),
such that $\alpha < 60^\circ$.  When $\alpha = 60^\circ$, $\epsilon=0$
and there is no shear---when both lattice parameters reach their
high-symmetry values, $\alpha = 60^\circ$ and $z=1/4$,
the structure becomes simple cubic.
The special \mbox{$\k$-points} of the rhombohedral BZ
indicated in Fig.~\ref{fig:a7bz2} have been written out explicitly
by Falicov and Golin,~\cite{falicovandgolin_65} who follow
Cohen's notation for points and lines of symmetry.~\cite{mhcohen_61}
These points of symmetry are expressed in fractional coordinates
(with respect to the reciprocal lattice vectors). Examples of each of these
points are:
\vspace{2mm}
\begin{equation}
 \label{eq:kpt_G}
   \Gamma = \left[ 0, 0, 0 \right],
\vspace{0.8mm}
\end{equation}
\begin{equation}
 \label{eq:kpt_X}
   \rm{X} = \left[ 0, \tfrac{1}{2}, \tfrac{1}{2} \right],
\vspace{0.8mm}
\end{equation}
\begin{equation}
 \label{eq:kpt_L}
   \rm{L} = \left[ 0, \tfrac{1}{2}, 0 \right],
\vspace{0.8mm}
\end{equation}
\begin{equation}
 \label{eq:kpt_T}
   \rm{T} = \left[ \tfrac{1}{2}, \tfrac{1}{2}, \tfrac{1}{2} \right],
\vspace{0.8mm}
\end{equation}
\begin{equation}
 \label{eq:kpt_W}
   \rm{W} = \left[ \gamma, 1-\gamma, \tfrac{1}{2} \right],
\vspace{0.8mm}
\end{equation}
\begin{equation}
 \label{eq:kpt_U}
   \rm{U} = \left[ \tfrac{1}{2} \gamma + \tfrac{1}{4}, 1-\gamma, \tfrac{1}{2} \gamma + \tfrac{1}{4} \right],
\vspace{0.8mm}
\end{equation}
\begin{equation}
 \label{eq:kpt_K}
   \rm{K} = \left[ 0, \tfrac{3}{4} - \tfrac{1}{2} \gamma, \tfrac{1}{2} \gamma + \tfrac{1}{4} \right],
\vspace{0.8mm}
\end{equation}
where~\cite{falicovandgolin_65}
\begin{equation}
 \label{eq:gamma}
  \gamma = \frac {(1+ \frac{1}{2} \epsilon ^2)}{(2 + \epsilon)^2}.
\vspace{0.8mm}
\end{equation}
Multiplicities and symmetry elements of these special
\mbox{$\k$-points} can be found in Cohen.~\cite{mhcohen_61}
Other group theoretical considerations can be found 
in Falicov and Golin.~\cite{falicovandgolin_65}
Note that the three special points W, U and K depend on $\gamma$,
and thus on the rhombohedral angle $\alpha$ of the real-space
primitive cell---their coordinates will therefore change as the 
pressure is increased.
When $\alpha=60^\circ$, $\epsilon=0$ and $\gamma=\frac{1}{4}$. 
Thus in the two-atom representation of sc arsenic:
\begin{equation}
 \label{eq:kpt_W_sc}
   \rm{W}_\rm{sc} = \left[ \tfrac{1}{4}, \tfrac{3}{4}, \tfrac{1}{2} \right],
\vspace{0.8mm}
\end{equation}
\begin{equation}
 \label{eq:kpt_U_sc}
   \rm{U}_\rm{sc} = \left[ \tfrac{3}{8}, \tfrac{3}{4}, \tfrac{3}{8} \right],
\vspace{0.8mm}
\end{equation}
\begin{equation}
 \label{eq:kpt_K_sc}
   \rm{K}_\rm{sc} = \left[ 0, \tfrac{5}{8}, \tfrac{3}{8} \right].
\vspace{0.8mm}
\end{equation}
\vspace{1.2mm}
As $\alpha \rightarrow 90^\circ$, $\epsilon \rightarrow -0.5$ and $\gamma=\frac{1}{2}$.
Thus in the two-atom representation of bcc arsenic,
the points W and U merge with T, and the point K merges with X:
\vspace{2mm}
\begin{equation}
 \label{eq:kpt_W_bcc}
   \rm{W}_\rm{bcc} = \rm{U}_\rm{bcc} = \rm{T} = \left[ \tfrac{1}{2}, \tfrac{1}{2}, \tfrac{1}{2} \right],
\vspace{0.8mm}
\end{equation}
\begin{equation}
 \label{eq:kpt_K_bcc}
   \rm{K}_\rm{bcc} = \rm{X} = \left[ 0, \tfrac{1}{2}, \tfrac{1}{2} \right].
\vspace{0.8mm}
\end{equation}
The convention is to label the three mutually orthogonal axes making up
the coordinate system of the A7 BZ as follows.
Referring again to Fig.~\ref{fig:a7bz2},
the binary axis is labeled the $x$-axis and contains the segment $\Gamma$K---it
is parallel to the TW line.  The bisectrix axis is 
labeled as the $y$-axis---it is parallel to the TU line but passes
through $\Gamma$.  The trigonal ($z$) axis passes through the segment $\Gamma$T.
The trigonal and bisectrix axes define a mirror plane.
Angles measured in the BZ are reported according to
a convention that is discussed in Refs.~\onlinecite{falicovandlin_66,
windmiller_66,priestley_67,ih_70} and in Ref.~\onlinecite{cooper_71}.
In the trigonal-bisectrix (mirror) plane, or some
parallel plane, angles are measured with respect to the 
vertical (the trigonal
axis, $\Gamma$T, or some parallel line).
Positive rotations are in the sense from $\Gamma$T to 
$\Gamma$X in the first quadrant of the coordinate system.
Our calculations of the unit cell of arsenic
yield the following angles for the special points in the
mirror plane:
$\Gamma$T is at $0^\circ$, $\Gamma$X is at $+58.4^\circ$,
$\Gamma$U is at $+41.3^\circ$ and $\Gamma$L is at $-72.9^\circ$.
These values are for the configuration
resulting from a geometry optimization of A7 arsenic at 0~GPa (and 0~K)
using the Perdew-Burke-Ernzerhof generalized gradient 
approximation~\cite{perdew_96} (abbreviated in this work as GGA-PBE)
for the exchange-correlation functional.
The details of these calculations can be found in Ref.~\onlinecite{silas_yates_haynes_08}.
We will make some brief comparisons here and there
between the two-atom and one-atom (primitive) cells
of sc arsenic.  The BZ
for the one-atom primitive sc structure
can be found for example in Ref.~\onlinecite{mhcohen_64}
or more recently in Ref.~\onlinecite{martin}.
Examples of the special points of this BZ
are $\Gamma=[0,0,0]$, R$=\left[\frac{1}{2},\frac{1}{2},\frac{1}{2}\right]$,
M$=\left[\frac{1}{2},\frac{1}{2},0 \right]$ and
X$=\left[\frac{1}{2},0,0\right]$.  The BZ of the 
primitive sc structure is the ``unfolded'' version
of the BZ of the two-atom unit cell in the sc
configuration.  In the two-atom BZ, the point R 
folds onto the point $\Gamma$, and the point M folds onto the 
point X.~\cite{mhcohen_64}
%

\section{Choice of the Number of Wannier Functions to Construct}
\label{sec:num_wfs}

%
\begin{figure}[htpb]
\centering
\includegraphics[width=\columnwidth]{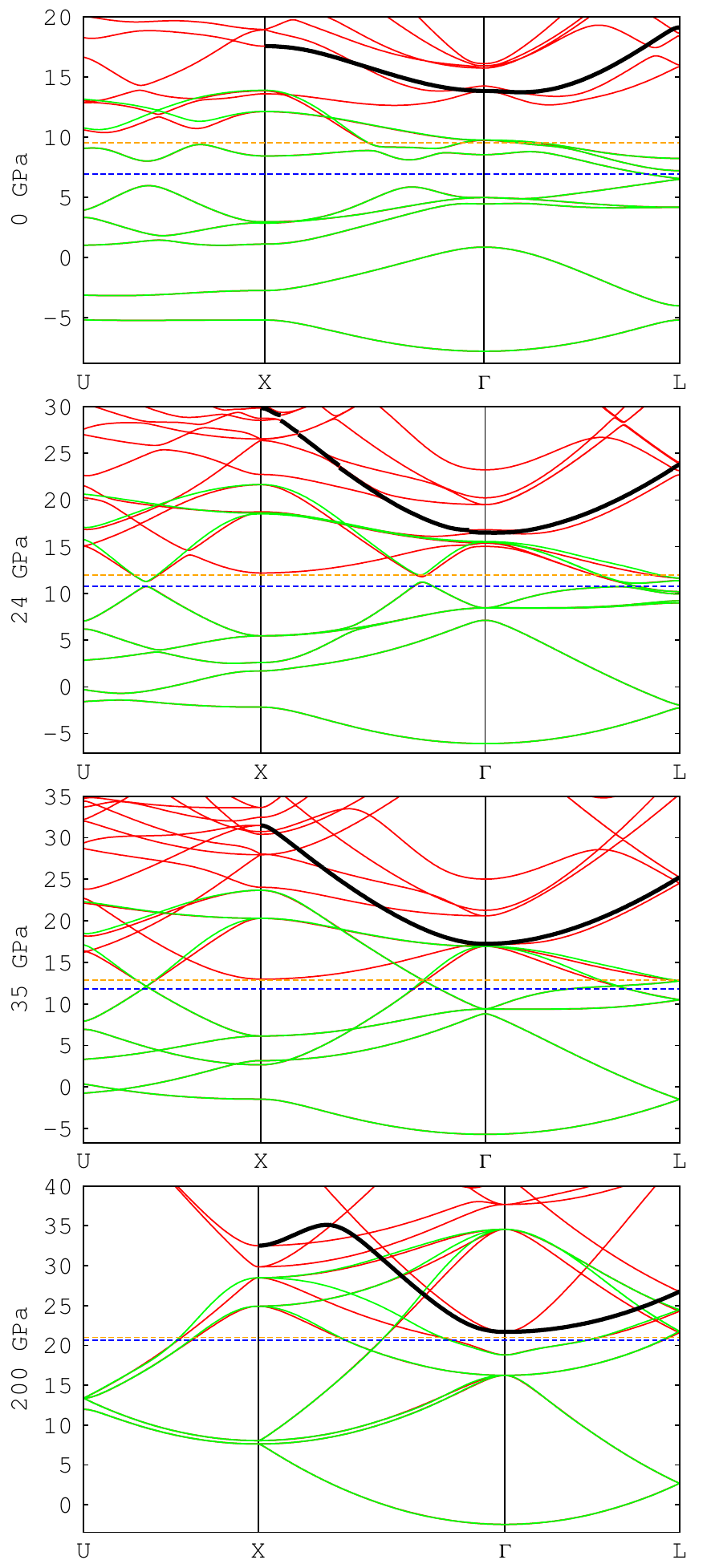}
\caption {\label{fig:bs_4wfs} \footnotesize Band structures of arsenic at
0, 24, 35 and 200~GPa (energy in~eV).  The Wannier-interpolated bands (in green)
have been obtained using four Wannier functions per atom of the unit cell.
The dotted blue line is the Fermi level and the dotted pink line indicates
the top of the ``inner window''.
Arsenic is in the A7 phase at 0 and 24~GPa, in the sc
phase at 35~GPa, and in the bcc phase at 200~GPa.  Using four
Wannier functions is not suitable for arsenic
at higher pressures.
As the pressure increases certain \textit{ab initio} bands can be seen to be
descending toward the Fermi level.
One of these states is outlined in black---this
state is not captured by the four $sp^3$ orbitals used as the initial
guess for the Wannier functions.}
\end{figure}
Initially, we began our studies using four Wannier functions
per atom of the unit cell---the initial guesses
for these were the hybrid $sp^3$ orbitals.~\cite{mostofi_08}
However as is demonstrated in
Fig.~\ref{fig:bs_4wfs}, in which the Wannier-interpolated
bands (in green) have been obtained using four Wannier functions per atom of the unit cell,
this choice turns out not to be well suited to studying
arsenic at higher pressures.  This is evidenced by the inability
of the four Wannier functions to capture the states of
the (red) \textit{ab initio} band structure
which can be seen to be descending (one of which in heavy black)
toward the Fermi level
with increasing pressure.
In addition, in order for the band structure to be faithfully
reproduced using only four Wannier functions at the
higher pressures, we would be obliged to keep the
``inner window'' quite close to the Fermi level
(see especially the bottom panel of Fig.~\ref{fig:bs_4wfs},
the band structure of arsenic at 200~GPa),
yet we would like ideally to be able to place it
a few eV above the Fermi level.
(The inner window is the energy window within which
the \textit{ab initio} band structure is required to be preserved
as per the disentanglement method of Souza, Marzari and Vanderbilt.~\cite{souza_01})
We choose therefore to address both points by working with
nine Wannier functions per
atom of the unit cell over the entire range of
pressures studied.  The trial functions used for the
initial guesses of the nine Wannier functions are
the six $sp^3 d^2$ hybrid orbitals in
addition to the $d_{xy}$, $d_{yz}$ and $d_{xz}$
orbitals,~\cite{mostofi_08} all atom-centered,
and this turns out to be entirely satisfactory.
%

\section{Further Details of the Electron and Hole Fermi Surfaces}
\label{sec:supp_results}

\subsection{Features of the Electron Fermi Surface}
\label{sec:el_fs_of_As}

%
\begin{figure}[t]
\centering
\includegraphics[width=\columnwidth]{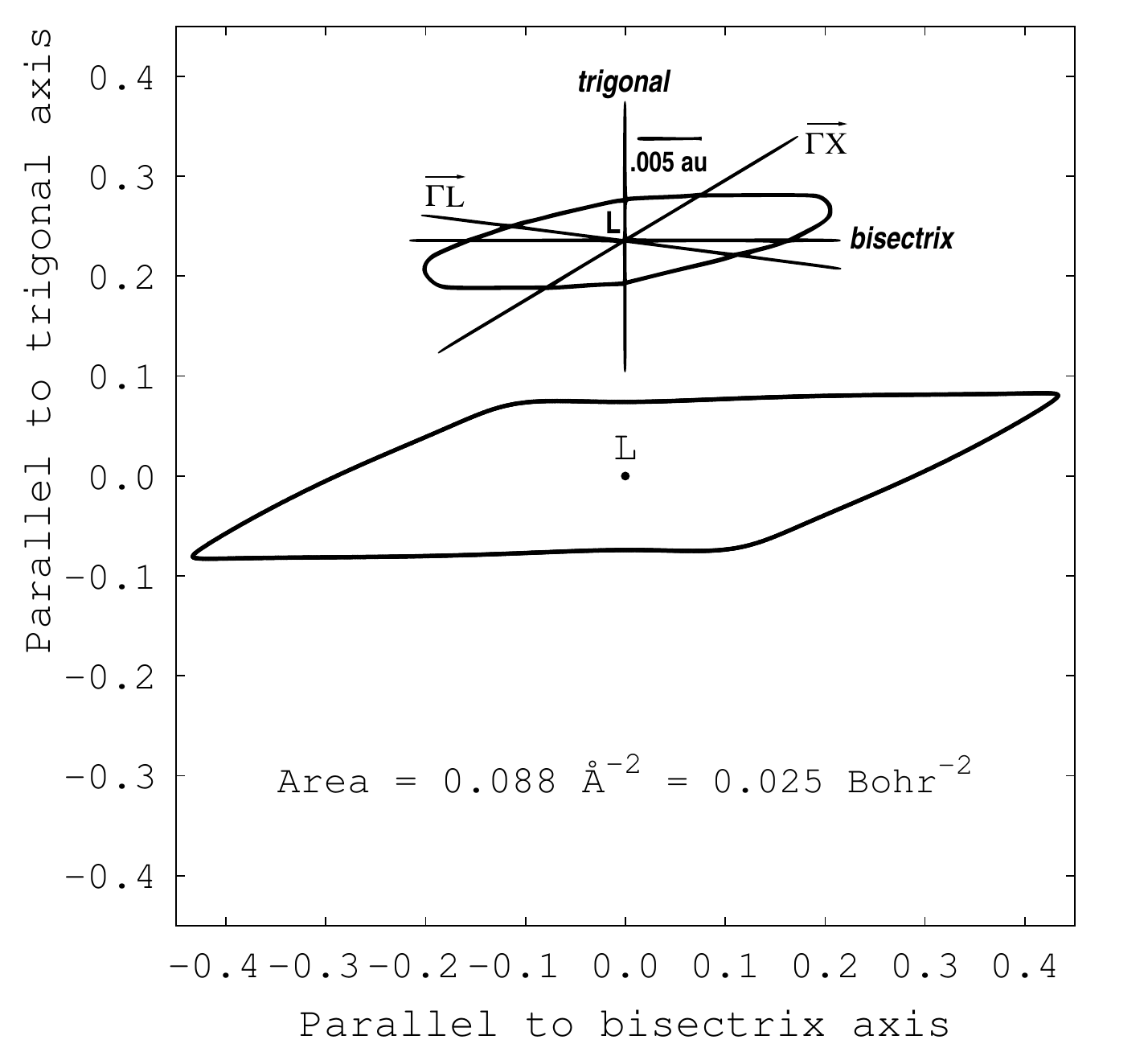}
\caption {\label{fig:electron_1} \footnotesize Electron pocket: trigonal-bisectrix plane through L.
This figure and the resulting area have been obtained
using a Fermi energy recomputed from the Wannier-interpolated
DOS of A7 arsenic at 0~GPa. 
All distances are in $\AA^{-1}$.
This contour can be located
in the top panel of the left-hand column of Fig.~6. The
inset is the analogous cross section as calculated by Lin and Falicov.~\cite{linandfalicov_66}}
\end{figure}
The electron Fermi surface of arsenic at 0~GPa is composed of three
ellipsoid-like pockets centered at the three equivalent L points
of the BZ.
The intersection of the trigonal-bisectrix plane
(the mirror plane) with an electron pocket at L results in the Fermi contour
illustrated in Fig.~\ref{fig:electron_1}.  We have included as an inset of
this figure the analogous cross section obtained by Lin and Falicov
in 1966.~\cite{linandfalicov_66}
\begin{figure}[htpb]
\centering
\includegraphics[width=\columnwidth]{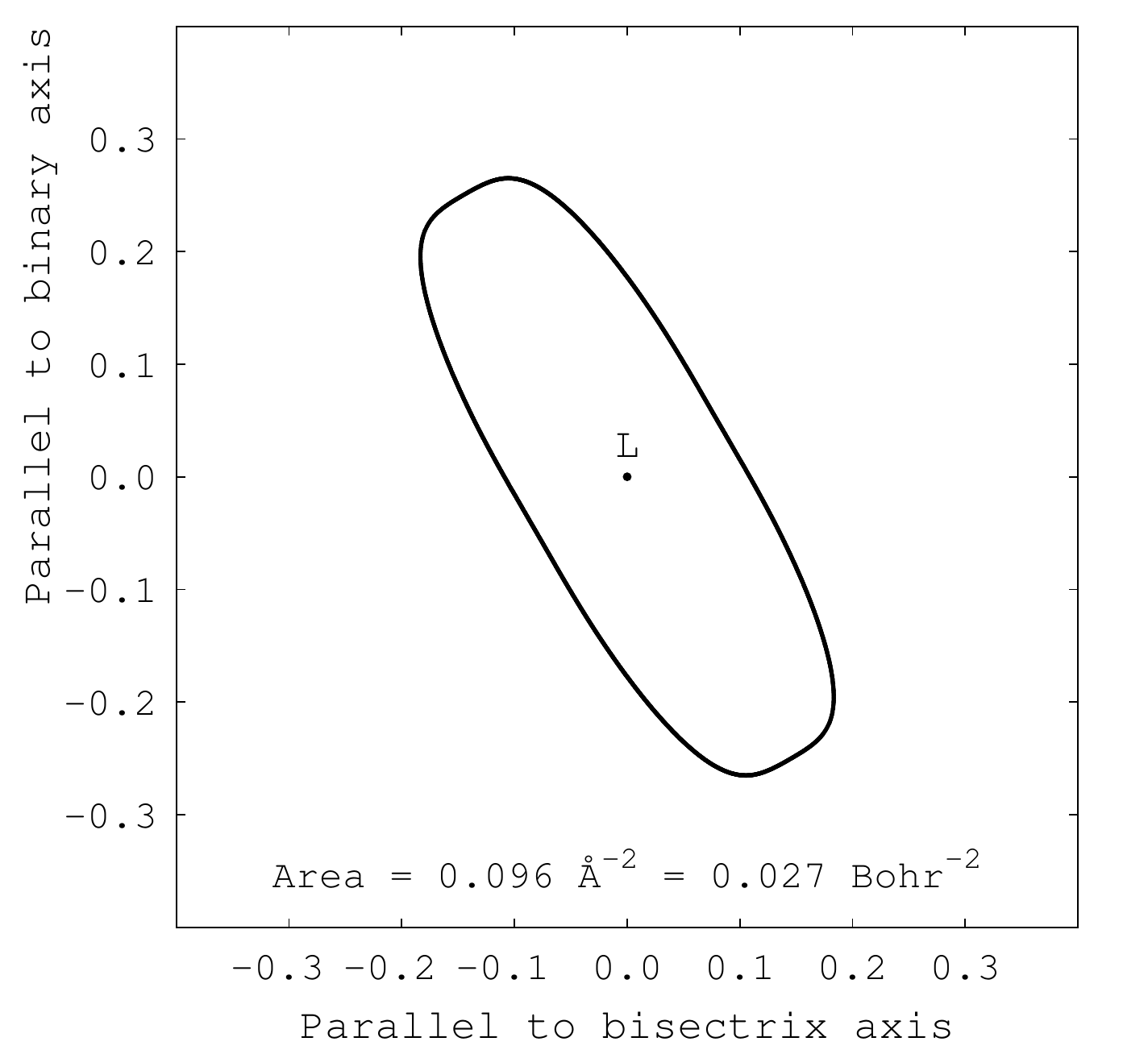}
\caption {\label{fig:electron_2} \footnotesize Electron pocket: binary-bisectrix plane through L.
This figure and the resulting area have been obtained
using a Fermi energy recomputed from the Wannier-interpolated
DOS of A7 arsenic at 0~GPa. 
All distances are in $\AA^{-1}$.
This contour can be located
in the top panel of the middle column of Fig.~6.} 
\end{figure}
Before we proceed, it is important to note that the Lin and Falicov calculations
were based on experimental data (from de Haas-van Alphen experiments performed
at temperatures between 1.2 and 4.2~K) available at the time---most of it
available to them by private communication,~\cite{linandfalicov_66} values
which were later somewhat altered upon their publication.~\cite{priestley_67} 
Lin and Falicov fixed the Fermi energies of the electron and hole Fermi
surfaces so as to fit the data available for certain features contained therein.
In the case of the electron Fermi surface, the Fermi energy was adjusted
to fit an area corresponding to the minimum area observed for magnetic fields
in the trigonal-bisectrix plane.~\cite{linandfalicov_66,priestley_67}
In the case of the hole Fermi surface, the Fermi
energy was adjusted to fit the area reported by Priestly, et al.~\cite{priestley_67}
to correspond to a cross section of one of the long thin necks.
Thus the calculations of Lin and Falicov have been adjusted to fit some prevalent
experimental data. Nevertheless it is still edifying to compare our results
with theirs.
Returning thus to Fig.~\ref{fig:electron_1}, we 
see that the contour of our electron pocket in the trigonal-bisectrix plane
is quite a bit sharper than that of Lin and Falicov.  
The area of this cross section,
0.088~\footnotesize$\AA^{-2}$\normalsize, is approximately $20\%$ larger than that obtained
from those early experiments, and it is approximately $55\%$ larger than that of Lin and
Falicov.  The long axis of the electron pocket makes an angle of $10.6^\circ$ with 
the horizontal---this is similar to what was found both theoretically~\cite{linandfalicov_66}
and experimentally.~\cite{priestley_67,cooper_71}
Careful inspection of Fig.~\ref{fig:electron_1} leads us to believe that
this cross section is somewhat ``S-shaped''.  This observation is corroborated
experimentally for arsenic by Priestley, \textit{et al.}~\cite{priestley_67} and
by Cooper and Lawson,~\cite{cooper_71} who performed cyclotron resonance experiments
for arsenic at 1.15~K.  This ``S-shape'' has also been calculated by Falicov and 
Lin as being the case for the analogous cross section
in antimony.~\cite{falicovandlin_66}
\begin{figure}[t]
\centering
\includegraphics[width=\columnwidth]{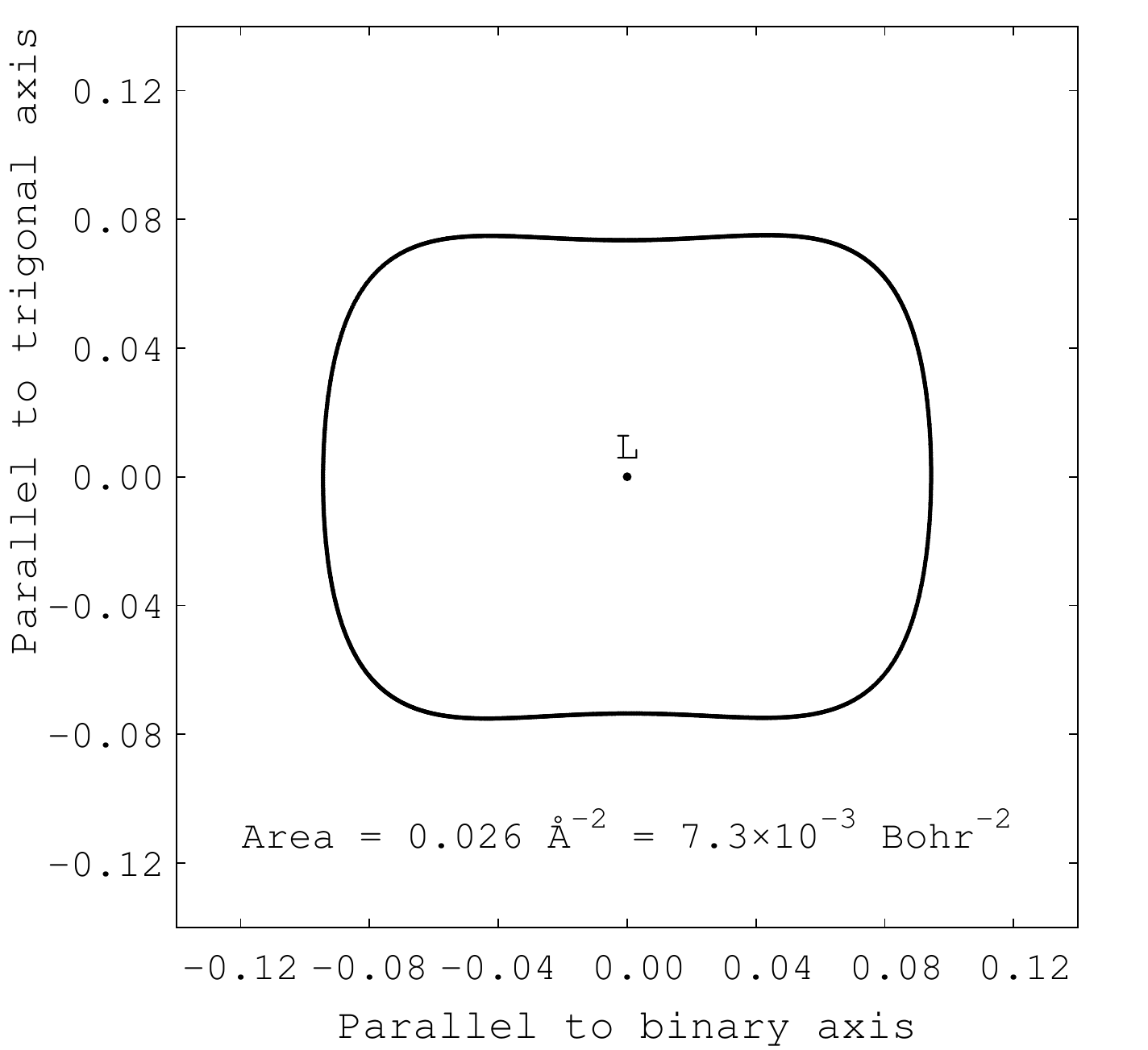}
\caption {\label{fig:electron_3} \footnotesize Electron pocket: trigonal-binary plane through L.
This figure and the resulting area have been obtained
using a Fermi energy recomputed from the Wannier-interpolated
DOS of A7 arsenic at 0~GPa.
All distances are in $\AA^{-1}$.
} 
\end{figure}

\begin{figure}[t]
\centering
\includegraphics[width=\columnwidth]{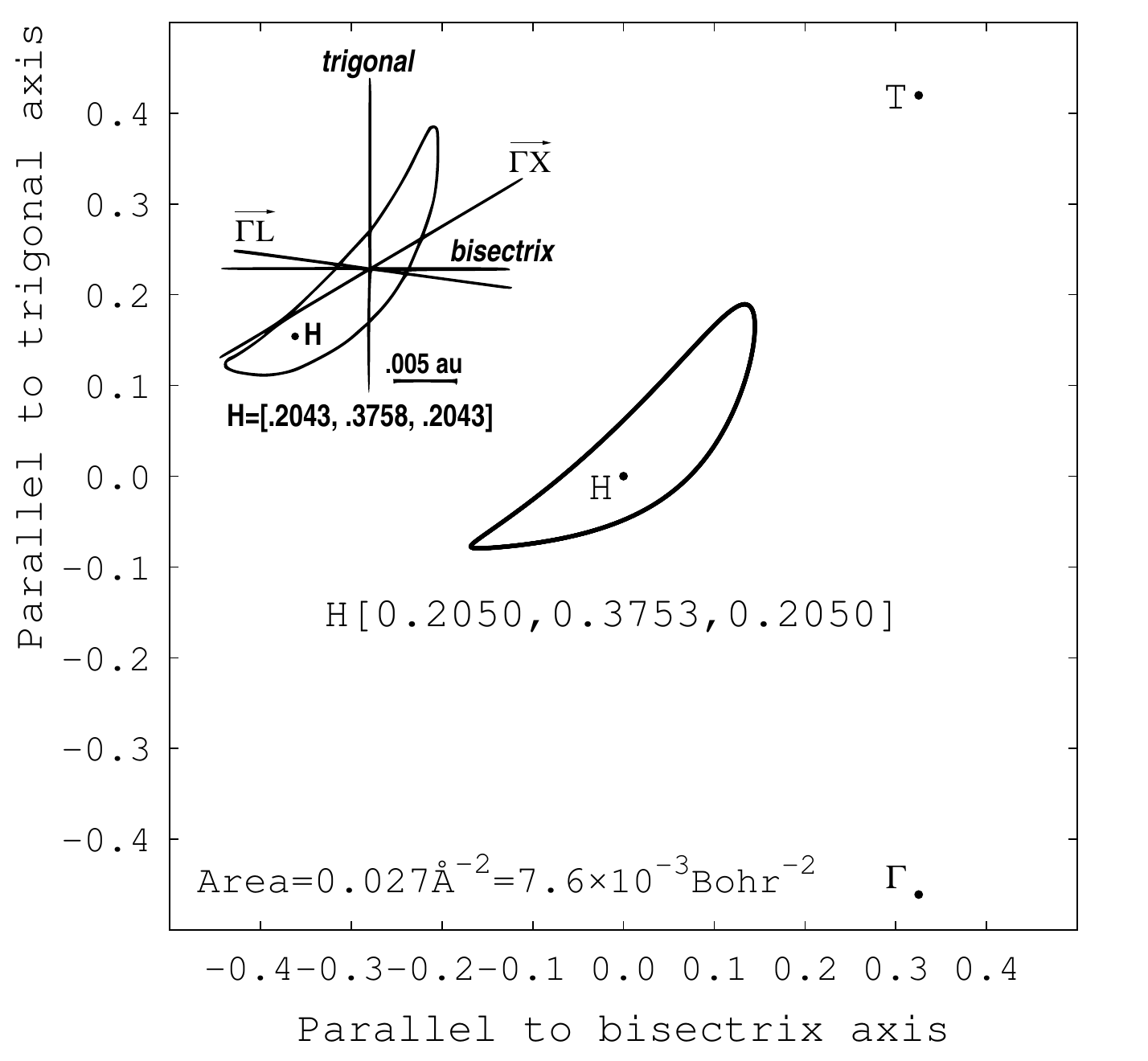}
\caption {\label{fig:hole_1} \footnotesize Hole pocket---one of six lobes of the hole crown:
 trigonal-bisectrix plane through H.
This figure and the resulting area have been obtained
using a Fermi energy recomputed from the Wannier-interpolated
DOS of A7 arsenic at 0~GPa. 
All distances are in $\AA^{-1}$.
This contour can be located
in the top panel of the left-hand column of Fig.~6.
The inset is the analogous cross section as
calculated by Lin and Falicov.~\cite{linandfalicov_66}  The points $\Gamma$ and T are
included as an aid to viewer orientation.}
\end{figure}
\begin{figure}[h]
\centering
\includegraphics[width=\columnwidth]{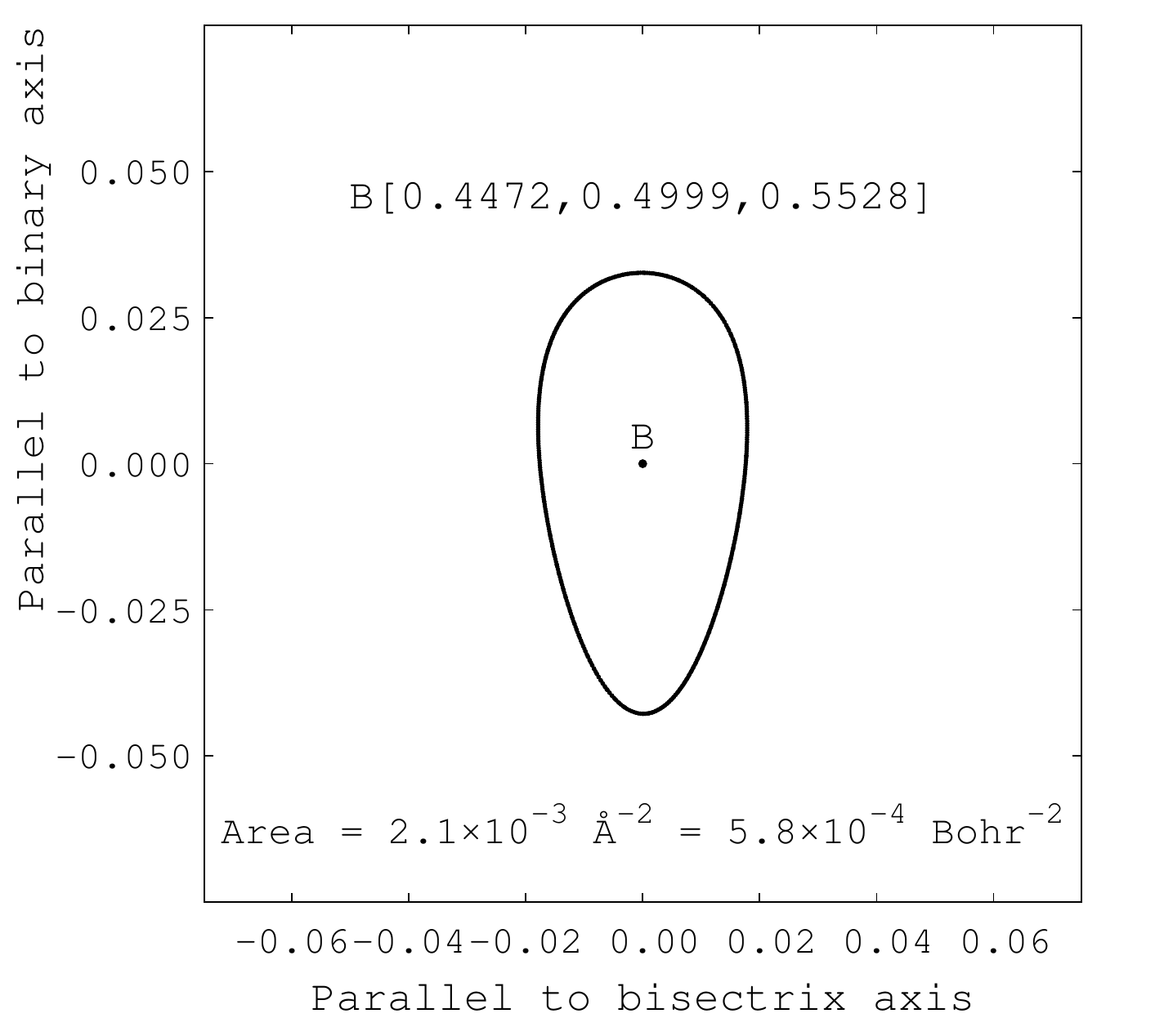}
\caption {\label{fig:hole_2} \footnotesize Hole pocket---one of six necks of the hole crown:
 binary-bisectrix plane through T.
This figure and the resulting area have been obtained
using a Fermi energy recomputed from the Wannier-interpolated
DOS of A7 arsenic at 0~GPa. 
All distances are in $\AA^{-1}$.
This contour can be located
in the top panel of the middle column of Fig.~6.}
\end{figure}
The intersection of the binary-bisectrix plane with an electron pocket at L
results in the cross section 
of area 0.096~\footnotesize$\AA^{-2}$\normalsize
presented in Fig.~\ref{fig:electron_2}, and the intersection of the trigonal-binary
plane with an electron pocket at L yields the contour
of area 0.026~\footnotesize$\AA^{-2}$\normalsize
displayed in Fig.~\ref{fig:electron_3}---this
cross section corresponds to that obtained by slicing the electron pocket
of Fig.~\ref{fig:electron_1} in the vertical direction at L.  These results,
as well as those that follow for the hole Fermi surface of arsenic at 0~GPa,
are summarized in Table~I, along with data resulting
from theory and experiment available in the literature.
\begin{figure}[h!]
\centering
\includegraphics[width=\columnwidth]{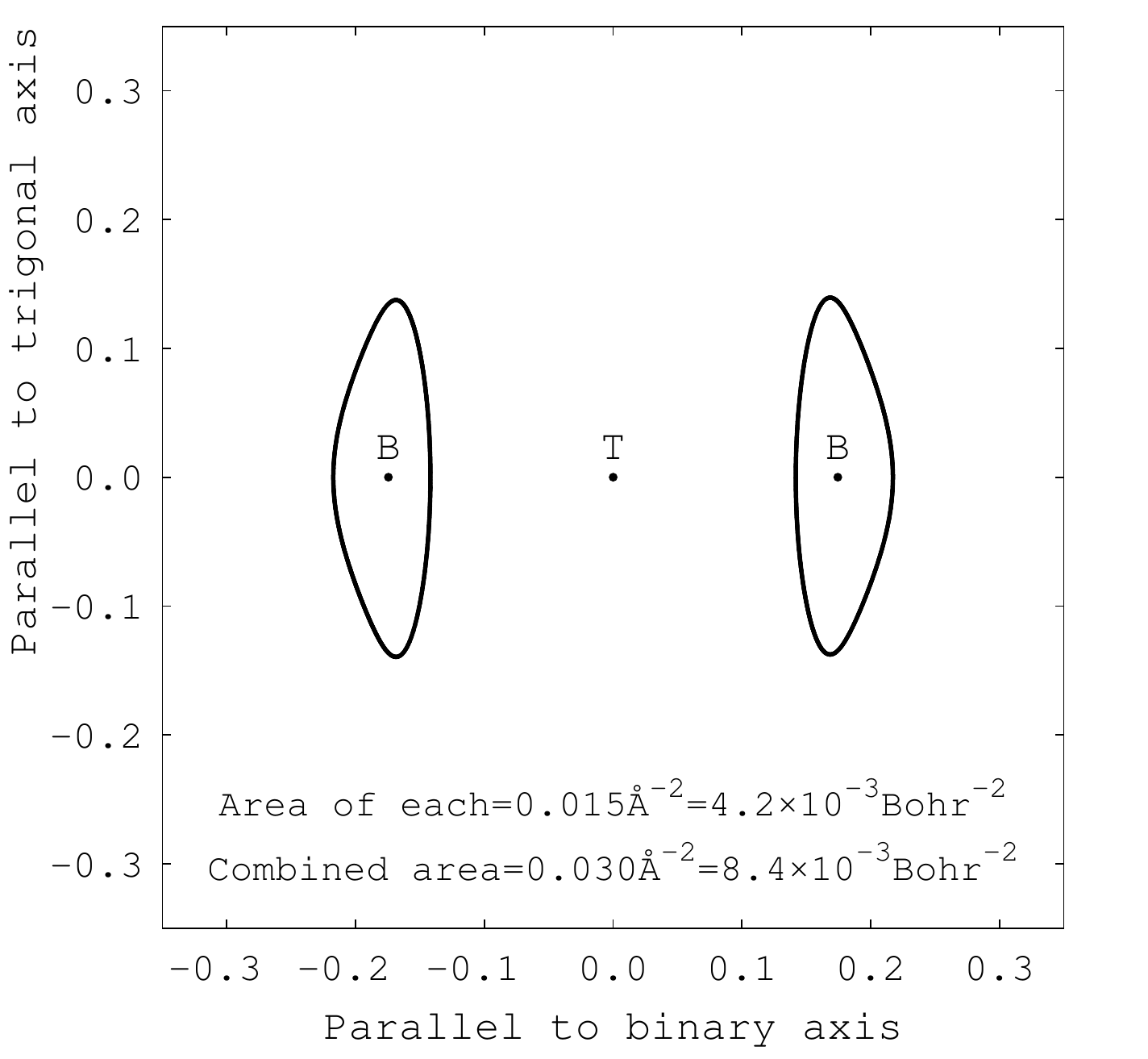}
\caption {\label{fig:hole_4} \footnotesize Hole pocket: trigonal-binary plane through T.
This figure and the resulting areas have been obtained
using a Fermi energy recomputed from the Wannier-interpolated
DOS of A7 arsenic at 0~GPa.
All distances are in $\AA^{-1}$.
}
\end{figure}
\begin{figure*}[t!]
\centering
\includegraphics[width=1.0\textwidth]{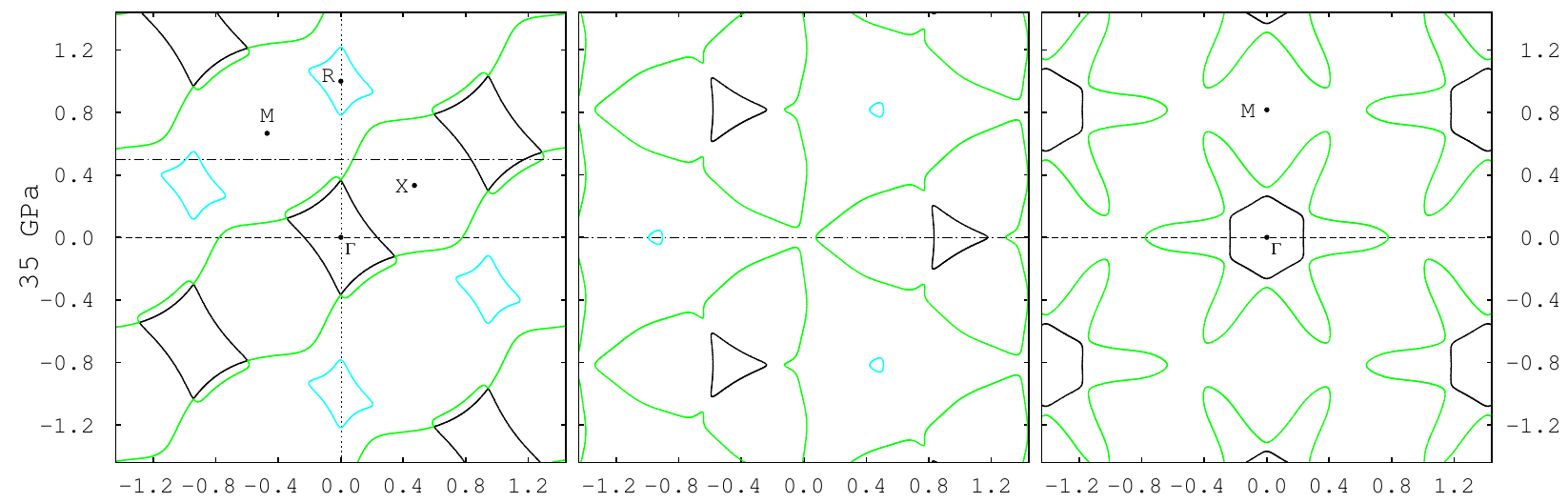}
\caption {\label{fig:35GPa_prim_cell} \footnotesize This figure is the unfolded version of the bottom
row of Fig.~6, obtained by slicing up the BZ
of the corresponding one-atom primitive cell of sc arsenic at 35~GPa.
The orientations of each of the three panels correspond to those of
Fig.~6, as do the dotted, dashed and dotted-dashed lines.
Distances in this figure are fractional with respect to the length of
the reciprocal lattice vector of the corresponding two-atom unit cell.
This is so that this figure can be overlayed and compared to the bottom
row of Fig.~6.  The second, third and fourth bands
contribute the black, green and blue contours, respectively.  In particular,
the second band provides cubic hole pockets at $\Gamma$ (in the left panel these are seen sliced
along the body diagonal) and the fourth band provides cubic electron pockets
at the points~R.}
\vspace{-3mm}
\end{figure*}

\subsection{Features of the Hole Fermi Surface}
\label{sec:hole_fs_of_As}

%
The hole Fermi surface of arsenic at 0~GPa is an object
centered at T composed of six lobe-like pockets (three up, three down)
connected by long thin cylinders or necks.
The intersection of the trigonal-bisectrix plane (the mirror plane)
with one of the lobe-like hole pockets results in the Fermi contour
displayed in Fig.~\ref{fig:hole_1}.  Once again, we include as an inset
of this figure the analogous cross section obtained by Lin and
Falicov in 1966.~\cite{linandfalicov_66}  Again, our respective contours
are similar, though slightly different toward the extremities.
The area of our cross section is 0.027~\footnotesize$\AA^{-2}$\normalsize,
which is about twice as large as in Ref.~\onlinecite{priestley_67} (experiment),
approximately $1\%$ larger than in Ref.~\onlinecite{cooper_71} (experiment),
and approximately $20\%$ smaller than that obtained by Lin and Falicov (although in
Ref.~\onlinecite{priestley_67} this value of Lin and Falicov is retracted
and it is claimed instead that the value is difficult to determine).
The maximum of the fifth band, designated as ``H'', occurs in the mirror
plane. It is labeled in Fig.~\ref{fig:hole_1}, along with the
points $\Gamma$ and T, which have been included in the diagram for orientation purposes.
Using a 2-D interpolation grid with a density of 5000 \mbox{$\k$-points}
per length of the reciprocal lattice vector, we determine the coordinates
of one of these six H points to be $[0.2050, 0.3753, 0.2050]$
(fractional coordinates with respect to the reciprocal lattice vectors).
The coordinates for the corresponding H point as found by Lin and Falicov
are $[0.2043, 0.3758, 0.2043]$ and so the agreement is excellent.  We
use our point H to determine the angle that the lobe makes with the
vertical to be $37.8^\circ$ (this is the angle between HT and $\Gamma$T),
which agrees closely with experiment.~\cite{priestley_67}
The intersection of the binary-bisectrix plane through T with the 
hole Fermi surface yields the contours of the cross sections of the six
necks, each located about a point B along a TW line.  One such cross section
is depicted in Fig.~\ref{fig:hole_2}. 
Again using a 2-D interpolation grid with a density of 5000 \mbox{$\k$-points}
per length of the reciprocal lattice vector,  we determine the fractional coordinates
of one of these six B points to be $[0.4472, 0.4999, 0.5528]$.
The coordinates for one of the B points as found by Lin and Falicov
are $[0.4617, 0.5, 0.5383]$, close to our own results.
Finally, as we did with the electron Fermi surface, we look at the intersection
of the trigonal-binary plane passing through $\Gamma$ with the hole Fermi surface,
which is illustrated in Fig.~\ref{fig:hole_4}.  We see that in this orientation,
two necks are sliced vertically---they are on opposite sides \mbox{of T}.
Once again, all results obtained for both the electron and hole Fermi
surfaces of arsenic at 0~GPa are summarized in Table~I,
and contrasted against results previously published in the literature.
We now proceed to investigate further the evolution of the Fermi surface 
of arsenic with pressure.
%

\section[Pressure Dependence of the Fermi Surface of Arsenic]
{Pressure Dependence of the Fermi Surface of Arsenic: 
A7~$\to$~SC Transition}
\label{sec:pressure_dependence_of_fs_of_As}

\subsection{SC Arsenic Using the One-Atom Primitive Cell}

%
In Fig.~\ref{fig:35GPa_prim_cell} we display contours of the
Fermi surfaces of sc arsenic at 35~GPa using
the one-atom primitive cell. (The self-consistent calculation
for the ground state potential of sc arsenic at 35~GPa
using the one-atom primitive cell is performed using a 40$\times$40$\times$40
Monkhorst-Pack grid of \mbox{$\k$-points}.)
This figure is the ``unfolded'' version
of the bottom row of Fig.~6,
where the calculations have been performed for sc arsenic
at 35~GPa using the two-atom unit cell. The orientation
of the panels correspond between the two figures. 
The contours displayed in Fig.~\ref{fig:35GPa_prim_cell}
are provided by bands 2 (black), 3 (green) and 4 (blue). 
The second band provides hole pockets at $\Gamma$, and the fourth band provides
electron pockets at R.  The distances in Fig.~\ref{fig:35GPa_prim_cell}
are fractional with respect to the length of the reciprocal lattice
vector of the two-atom unit cell of sc arsenic---this
is so that the folded and unfolded versions of the figure
can be overlayed and compared with each other.  Points R and M that
appear in the left panel of Fig.~\ref{fig:35GPa_prim_cell}
fold onto the points $\Gamma$ and X respectively when the two-atom
unit cell is used.
The origin of the middle panel of the figure is the point
[$\tfrac{1}{4},\tfrac{1}{4},\tfrac{1}{4}$], corresponding
to the T point in the case of the two-atom unit cell---in order for the
middle panel of this figure to correspond to the middle panel
of the bottom row of Fig.~6, the plane 
containing the points T, W and U in the two-atom case is
slid downward by the fractional coordinates
[$\tfrac{1}{4},\tfrac{1}{4},\tfrac{1}{4}$].
\begin{figure}[htpb]
\centering
\includegraphics[width=\columnwidth]{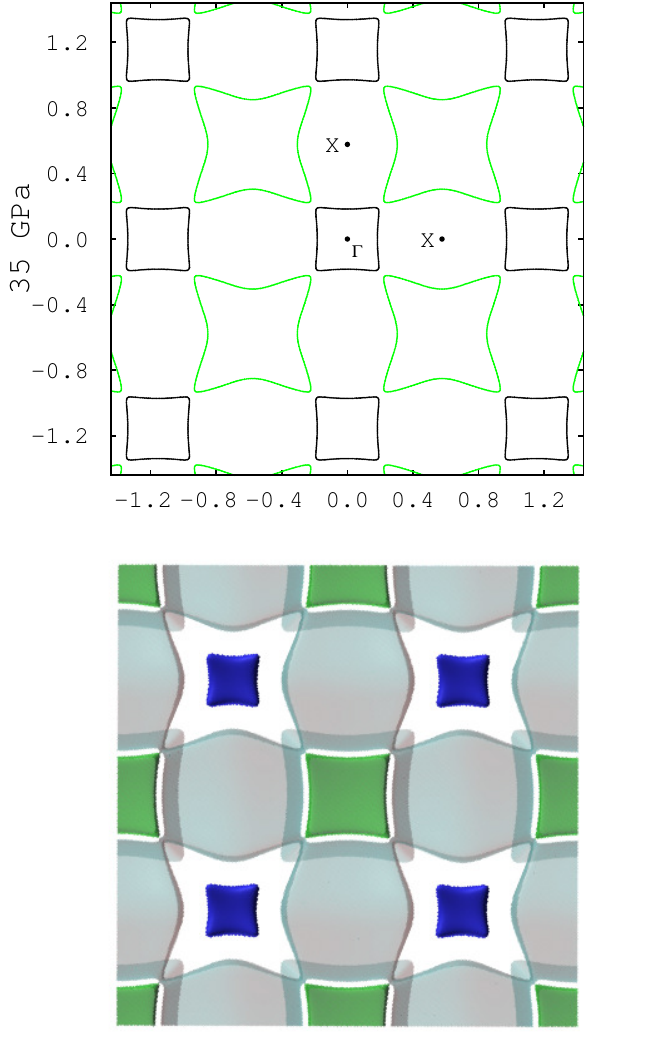}
\caption {\label{fig:35GPa_prim_cell_3rd_dir_and_FS} \footnotesize Another perspective of the Fermi
surface of sc arsenic at 35~GPa using the one-atom primitive cell.  In the top panel,
we see that for this slice through the BZ only the second and third
bands provide contours (in black and green, respectively).  This
perpective allows us to see the actual cubic cross section of the unfolded hole
pocket at $\Gamma$. Distances in the top panel are once again fractional
with respect to the length of the reciprocal lattice vector of the two-atom unit cell.
In the bottom panel,
the 3-D Fermi surface in the direction corresponding to the slice of the top panel is plotted.
The second band provides the dark green cube at $\Gamma$, the fourth band
provides the dark blue cubes at the points R, and the third band provides
the remainder of the object shown.}
\end{figure}
\begin{figure}[t]
\centering
\includegraphics[width=\columnwidth]{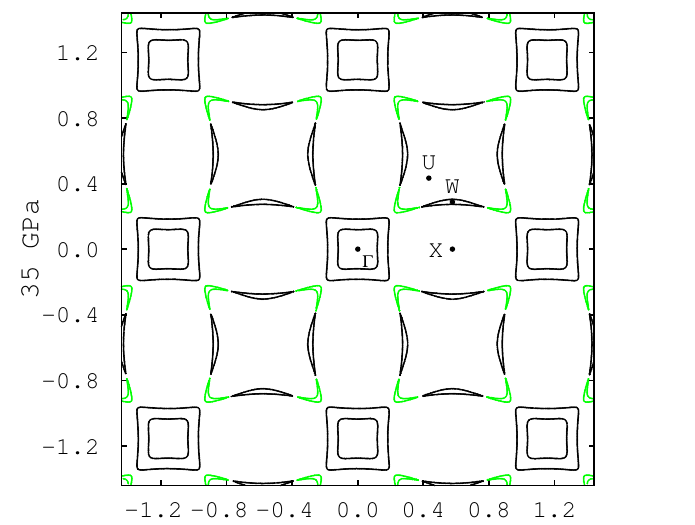}
\caption {\label{fig:35GPa_3rd_dir} \footnotesize Simple-cubic arsenic at 35~GPa using the two-atom unit cell:
this figure is the folded version of the top panel of Fig.~\ref{fig:35GPa_prim_cell_3rd_dir_and_FS}.}
\end{figure}
The top panel of Fig.~\ref{fig:35GPa_prim_cell_3rd_dir_and_FS} depicts
another view of the unfolded Fermi surface of arsenic in the sc
phase at 35~GPa.  This slice through the one-atom sc 
BZ contains $\Gamma$ and the points X, and we can see
the cubic hole surface centered on $\Gamma$ provided by the second band.
The green contours are provided by the third band.  The fourth band does not cross
the Fermi level for any of the \mbox{$\k$-points} on this slice
through the sc BZ, hence the fourth band
provides no contours to this slice.  
In order to be consistent, distances once again are with
respect to the length of the reciprocal lattice vector of the 
two-atom unit cell.
In the bottom panel of Fig.~\ref{fig:35GPa_prim_cell_3rd_dir_and_FS},
the \texttt{XCRYSDEN} package~\cite{xcrysden} has been used to plot
out the three-dimensional Fermi surface corresponding to the view
displayed in the top panel of the figure.  Surfaces due
to bands 2, 3 and 4 are merged to form the object shown.
The second band provides the dark green cube centered on $\Gamma$. 
The fourth band provides the dark blue cubes centered on R.
The corresponding slice of the BZ for the two-atom
representation of sc arsenic at 35~GPa is shown
in Fig.~\ref{fig:35GPa_3rd_dir}---it is the folded version
of the top panel of Fig.~\ref{fig:35GPa_prim_cell_3rd_dir_and_FS}.
%

\subsection{Higher-Pressure Transitions of Arsenic}

%
\begin{figure*}[htpb]
\centering
\includegraphics[width=1.0\textwidth]{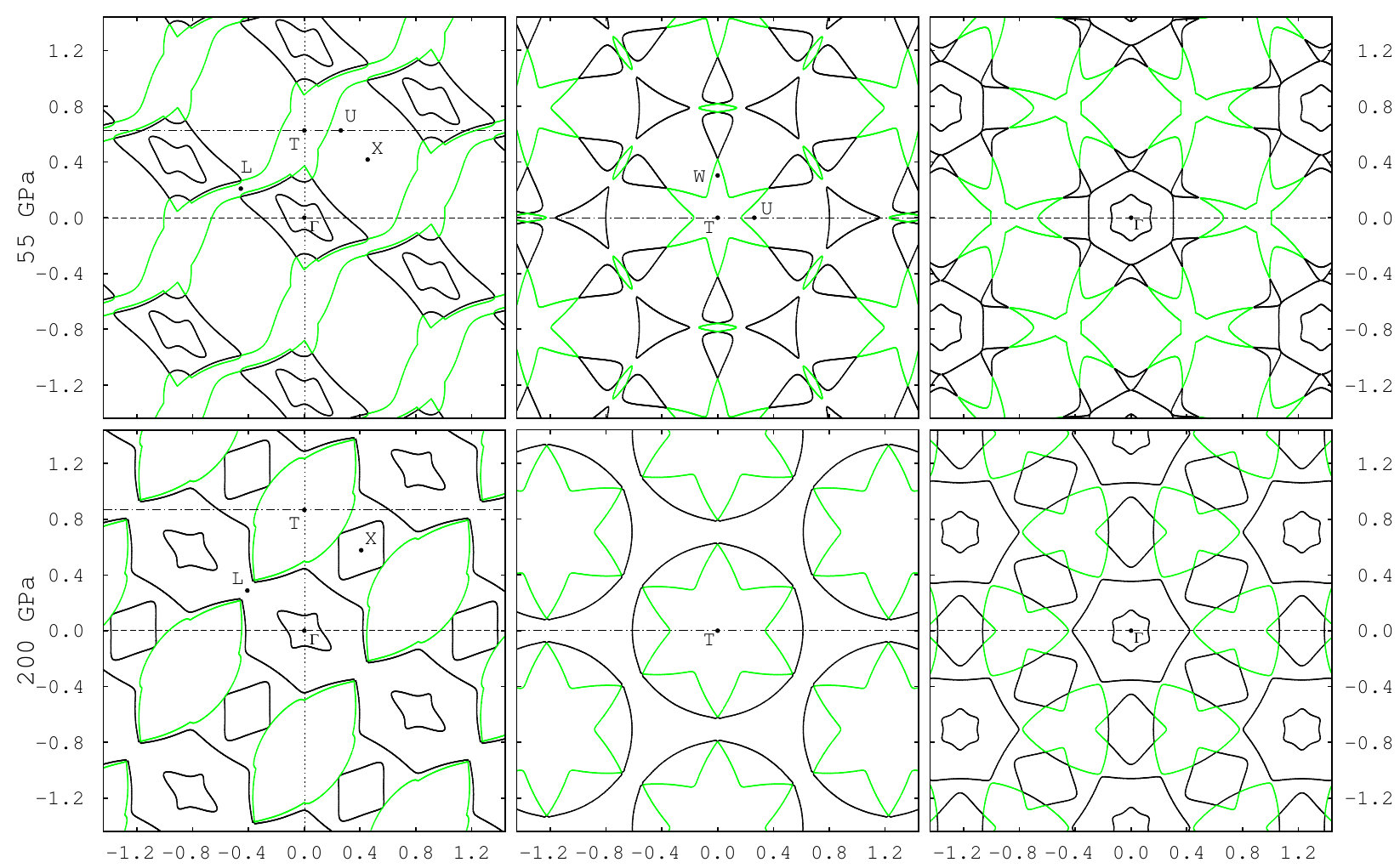}
\caption {\label{fig:55-200GPa} \footnotesize Pressure dependence of the Fermi surface of arsenic: \mbox{55--200GPa}.
This figure is in the style of Fig.~6.  The top row represents slices of the BZ
for arsenic at 55~GPa, when it is in our fictitious ``incomm'' phase. In the bottom row
we display the slices for bcc arsenic at 200~GPa.}
\end{figure*}

Watching Animations~01--03,\cite{animation1,animation2,animation3}
we observe that arsenic ceases to be in the sc phase
over 53--54~GPa, and enters the bcc phase over
95--96~GPa---there are quite dramatic changes to
the Fermi contours taking place at these transition pressures.
The pressures at which these transitions occur are as
expected since this work is based on our earlier findings,\cite{silas_yates_haynes_08}
as we have already mentioned. 
Between the sc and bcc phases,
arsenic has an incommensurate structure (with increasing
pressure, the progression of the phases of arsenic is
A7~$\to$~sc~$\to$~incomm~$\to$~bcc
as discussed in Ref.~\onlinecite{silas_yates_haynes_08}).  As stated
in that work, we do not attempt to model properly
the phase existing between sc and bcc---a
periodically-repeated two-atom unit cell cannot model 
an incommensurate structure. We have acknowledged this point
by simply labeling as ``incomm'' the structure we find between
the sc and bcc phases.
That being stated, in the interest of thoroughness and continuity we nevertheless
use our earlier findings for this ``incomm'' phase (over the range
of 54--95~GPa) in the same way as we have done for the other pressures studied.
At 55~GPa, arsenic is in our fictitious ``incomm'' phase. At 200~GPa,
arsenic is in the bcc phase.
In Fig.~\ref{fig:55-200GPa} we display for 55~GPa and 200~GPa the 
left-hand, middle and right-hand columns in the style of 
Fig.~6: the panels of the left-hand column
refer to intersections of the trigonal-bisectrix plane with
the BZ, those of the middle column
to intersections of the binary-bisectrix plane through T
with the BZ, and those
of the right-hand column to intersections
of the binary-bisectrix plane through $\Gamma$ with
the BZ.  Once again, all distances in 
this figure are fractional with respect to the length of
the reciprocal lattice vector.  The dotted, dashed and
dotted-dashed lines are as described for Fig.~6.
In the two-atom representation of the bcc phase
of arsenic, the points T, W and U merge.  This is evinced
by comparing both rows of Fig.~\ref{fig:55-200GPa}:
in the left-hand panels the points T and U are seen to be
merged at 200~GPa, while in the middle panels this is the 
case for all three points.
Only bands 5 (in black) and 6 (in green) are represented
in this figure, though at 55~GPa \mbox{bands 5--8} cross
the Fermi level and at 200~GPa it is crossed
by \mbox{bands 3--8}.  For our two-atom periodically-repeated
representation of arsenic at 55~GPa, \mbox{bands 7 and 8} missing
from the figure both provide small electron pockets at T.
In the 200~GPa case, missing \mbox{bands 3 and 4} give rise to
very small hole pockets at X, while \mbox{bands 7 and 8}
produce electron pockets at T.
In a further elucidation of the might of Wannier interpolation,
we conclude this section by inviting the reader to 
peruse some or all of the remaining animations
that accompany this work.~\cite{animation4,animation5,animation6,animation7,
animation8,animation9,animation10,animation11,animation12,
animation13,animation14,animation15,animation16} 
Of the 16 animations
that have been included, we have been discussing the
three which illustrate the evolution of the Fermi surface
of arsenic with pressure.~\cite{animation1,animation2,animation3}
For each of the six pressures (0, 10, 20, 35, 55 and 200~GPa)
that contribute to Fig.~6 and to Fig.~\ref{fig:55-200GPa},
we have plotted the two-dimensional band structure of the fifth
and sixth bands corresponding to a sequence of slices through
the BZ.  We slice the BZ in two different
directions producing two animations for each
pressure:~\cite{animation4,animation5,animation6,animation7,
animation8,animation9,animation10,animation11,animation12,
animation13,animation14,animation15}
parallel to the trigonal-bisectrix plane and parallel to the binary-bisectrix
plane.  In addition to the 2-D band structures, the resulting
electron and hole Fermi contours are also
plotted.
Lastly, as we have also been working with the one-atom
primitive cell of arsenic at 35~GPa, we have created an
animation displaying the Fermi contours 
provided by bands 2, 3 and 4 resulting again from
such slices through the BZ.~\cite{animation16}  
The left-hand panel of each frame of this animation displays the contours
arising from a slice that is parallel to a plane
corresponding to the trigonal-bisectrix plane.
The right-hand panel displays the contours
resulting from slices parallel to a plane
corresponding to the binary-bisectrix plane in 
the two-atom case.
%

\section{Wannier-Interpolated Densities of States 
\mbox{of A7, SC and BCC Arsenic}}
\label{sec:DOS}

\begin{figure*}[htpb]
\centering
\includegraphics[width=1.0\textwidth]{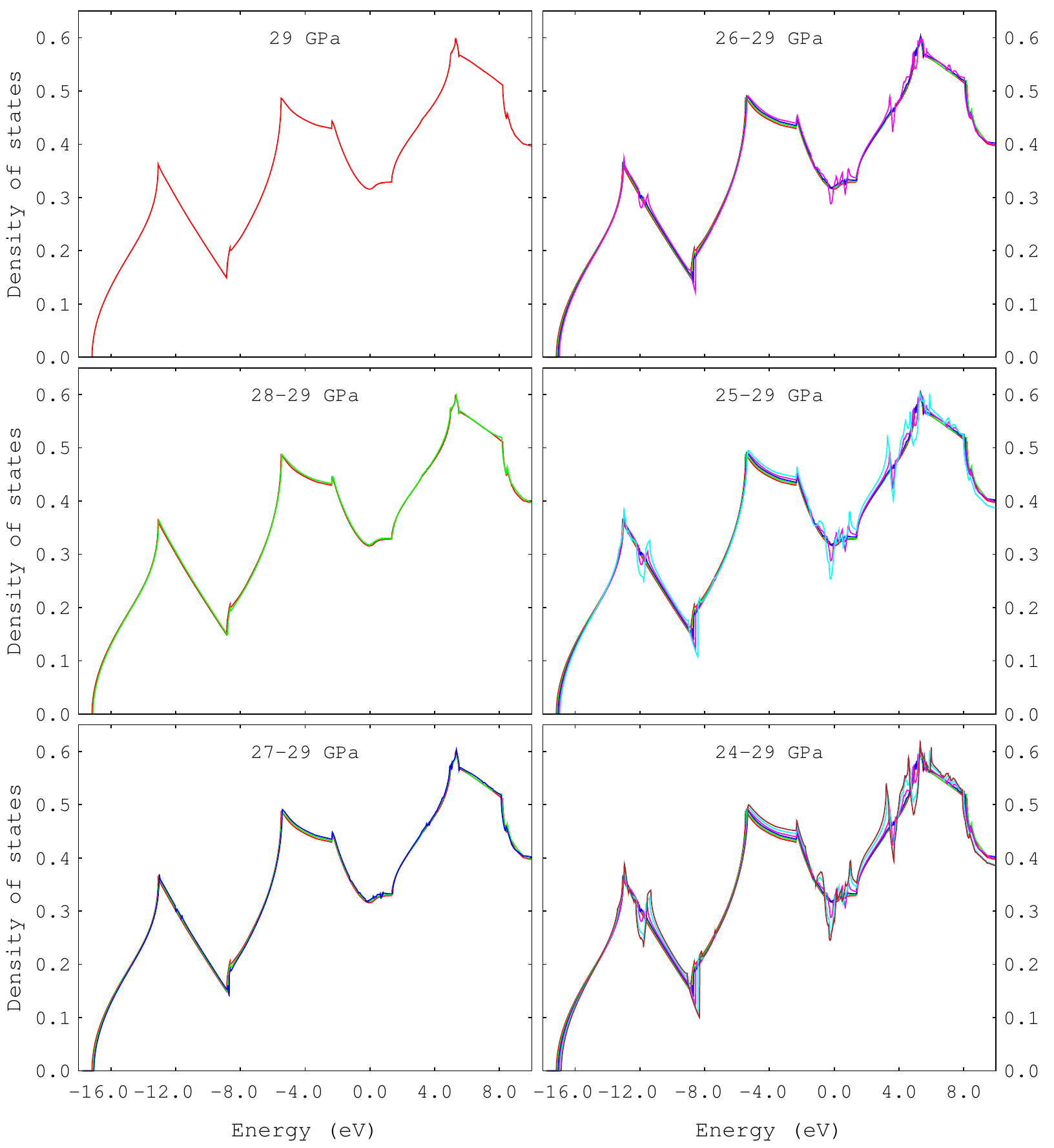}
\caption[\small Evolution of the DOS of arsenic across the sc~$\to$~A7 phase transition]
{\label{fig:DOS-29-24GPa} \footnotesize Evolution of the DOS of arsenic
across the sc~$\to$~A7 phase transition. The DOSs are referenced
to the Fermi level.  The panels are ordered from top to bottom 
starting with the left-hand column. The DOS of sc arsenic at 29~GPa is presented in
the top left panel.  DOSs at 1~GPa intervals and with decreasing pressure
are added to each successive image in the panels that follow. 
A marked change occurs between 26 and 27~GPa, as can be seen in the top right panel.
The emerging van Hove singularities indicate the onset of
 the Peierls-type cubic to rhombohedral distortion.
This agrees with the pressure at which the folding of the Fermi
surfaces is seen to occur, as discussed in Sec.~V and
which can be observed by viewing the relevant
animations.~\cite{animation1,animation2,animation3}
In Ref.~\onlinecite{silas_yates_haynes_08} we reported that the electronic change
that occurs across the transition drives the atomic positional
parameter $z$ to its high-symmetry value---our results here suggest instead
that it is the rhombohedral angle $\alpha$ that is being driven 
to its high-symmetry value.
}
\end{figure*}

\begin{figure*}[htpb]
\centering
\includegraphics[width=0.9\textwidth]{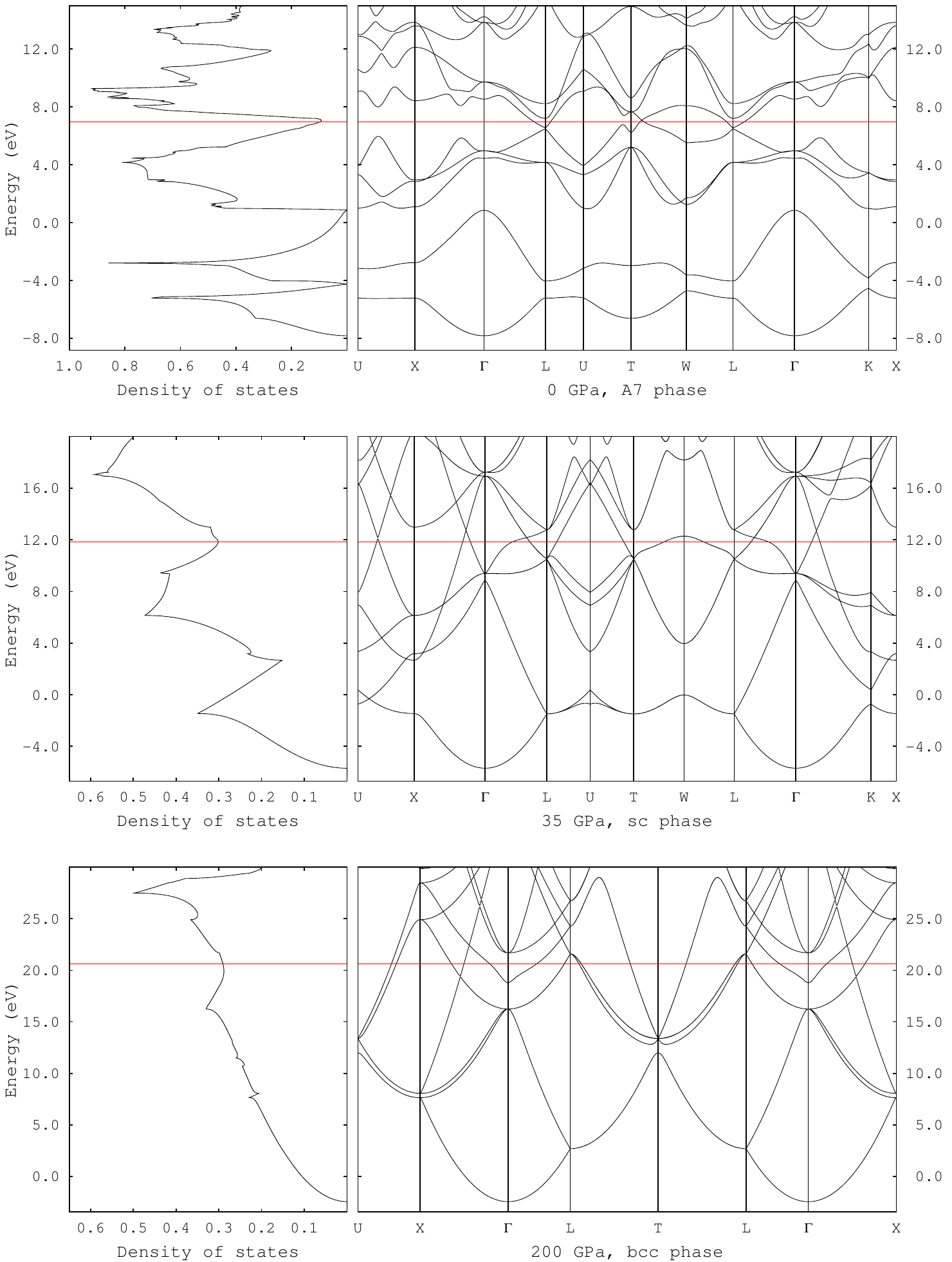}
\caption[\small The Wannier-interpolated DOSs of A7, sc and bcc arsenic]
{\label{fig:DOS-0-35-and-200GPa} \footnotesize  The Wannier-interpolated DOSs
of A7, sc and bcc arsenic and corresponding band structures.  In each case, the Fermi level
is indicated by the solid red line. 
At 0~GPa, arsenic is a semimetal---from the band structure
in the top panel we see that the fifth band provides tiny hole pockets
at a point (called B) close to T along the TW line.
The sixth band provides electron pockets
at L.  At 35~GPa, arsenic is a metal---the band structure displayed
for sc arsenic in the middle panel is ``folded''---it results
from performing our calculations on a unit cell containing two atoms.
Scanning the band structures from top to bottom, it can be seen
that low-lying bands appear free-electron-like with
increasing pressure.
}
\end{figure*}

%
Finally we inspect the evolution of the DOS
of arsenic in the immediate
vicinity of the A7~$\to$~sc transition, and present our results
in Fig.~\ref{fig:DOS-29-24GPa}.
This figure consists of a series of six panels that are ordered from 
top to bottom, starting with the left-hand column. 
The top left panel displays the DOS of arsenic at 29~GPa, when
it is in the sc phase.
In the panels that follow, we see
successive superpositions of the DOSs that result as the
pressure is decreased at intervals of 1~GPa.
In the bottom right
panel, the six DOSs from 29 to 24~GPa are shown.
Changes in the DOS through the sc~$\to$~A7 transition as the pressure
is decreased are more clearly seen in this direction than in the reverse.
We observe a marked change in the DOS between 26 and 27~GPa, as evinced
by the sharp wiggles that first appear in the top right panel of the figure.
These emerging van Hove singularities indicate the onset of the Peierls-type
cubic to rhombohedral distortion.
We reported in Ref.~\onlinecite{silas_yates_haynes_08} that in the GGA-PBE case
the A7~$\to$~sc transition occurs at 28$\pm$1~GPa, coinciding with the pressure
at which the atomic positional parameter $z$ reaches
its high-symmetry value of $1/4$.
The signature we notice in the top right panel of  Fig.~\ref{fig:DOS-29-24GPa}
however seems to coincide with the change in the cell angle $\alpha$ 
(we showed in Ref.~\onlinecite{silas_yates_haynes_08} that $\alpha$ reaches its
high-symmetry value at a lower pressure than does $z$).
In other words, the electronic change occurring as the pressure is increased
appears to be driving $\alpha$ to $60^\circ$.
Yet arsenic cannot be said to be in the sc phase until both $\alpha$ and $z$ 
have reached their high-symmetry values.  As reported
in Ref.~\onlinecite{silas_yates_haynes_08}, for GGA-PBE the difference between
these two pressures is approximately 2~GPa.
The above observation does coincide however with what we see in the animations
that accompany this work.~\cite{animation1,animation2,animation3}
As we discussed in Sec.~V,
a folding of the Fermi surfaces is seen to occur by 27~GPa.
It is useful to be able to compare DOS plots with the band structures
to which they correspond.
We offer in Fig.~\ref{fig:DOS-0-35-and-200GPa} just such a comparison
for the A7, sc and bcc phases of arsenic at
0, 35 and 200~GPa, respectively.  In each case, the Fermi level is indicated by
the solid red line.  As the pressure is increased,
the lowest bands begin to appear free-electron-like.  
%

\bibliographystyle{apsrev}

\end{document}